\documentclass[reprint,prl,aps, floatfix]{revtex4-1}
\usepackage{graphicx}
\usepackage{xcolor}
\usepackage{dcolumn}
\usepackage{bm}
\usepackage{amssymb}
\usepackage{braket}
\usepackage{color,amsmath}
\usepackage{comment}
\usepackage{gensymb}
\usepackage{soul,xcolor} 
\setstcolor{red} 
\usepackage{epstopdf}
\usepackage{hhline}
\usepackage{tabularx}
\usepackage{xspace}
\usepackage{layouts}
\usepackage{lineno} 
\usepackage{xr} 

\newcolumntype{C}[1]{>{\centering\arraybackslash}m{#1}}
\newcolumntype{N}{@{}m{0pt}@{}}

\newcommand{\moire}{moir\'e\xspace}

\begin{document}

\title{Interplay of electronic crystals with integer and fractional Chern insulators in moir\'e pentalayer graphene}

\author{Dacen Waters$^{1,2*}$}
\author{Anna Okounkova$^{1*}$}
\author{Ruiheng Su$^{3,4}$}
\author{Boran Zhou$^{5}$}
\author{Jiang Yao$^{1}$}
\author{Kenji Watanabe$^{6}$}
\author{Takashi Taniguchi$^{7}$}
\author{Xiaodong Xu$^{1,8}$}
\author{Ya-Hui Zhang$^{5}$}
\author{Joshua Folk$^{3,4}$}
\author{Matthew Yankowitz$^{1,8\dagger}$}

\affiliation{$^{1}$Department of Physics, University of Washington, Seattle, Washington, 98195, USA}
\affiliation{$^{2}$Intelligence Community Postdoctoral Research Fellowship Program, University of Washington, Seattle, Washington, 98195, USA}
\affiliation{$^{3}$Quantum Matter Institute, University of British Columbia, Vancouver, British Columbia, V6T 1Z1, Canada}
\affiliation{$^{4}$Department of Physics and Astronomy, University of British Columbia, Vancouver, British Columbia, V6T 1Z1, Canada}
\affiliation{$^{5}$Department of Physics and Astronomy, Johns Hopkins University, Baltimore, Maryland, 21205, USA}
\affiliation{$^{6}$Research Center for Electronic and Optical Materials, National Institute for Materials Science, 1-1 Namiki, Tsukuba 305-0044, Japan}
\affiliation{$^{7}$Research Center for Materials Nanoarchitectonics, National Institute for Materials Science, 1-1 Namiki, Tsukuba 305-0044, Japan}
\affiliation{$^{8}$Department of Materials Science and Engineering, University of Washington, Seattle, Washington, 98195, USA}
\affiliation{$^{*}$These authors contributed equally to this work.}
\affiliation{$^{\dagger}$myank@uw.edu (M.Y.)}

\maketitle

\textbf{The rapid development of moir\'e quantum matter has recently led to the remarkable discovery of the fractional quantum anomalous Hall effect~\cite{Cai2023,Zeng2023,Park2023,Xu2023fqah,Lu2024,Xie2024FQAH}, and sparked predictions of other novel correlation-driven topological states~\cite{Goldman2023cfl,Dong2023cfl,Reddy2023cfl,Xu2024ed,Wang2024ed,Chen2024ed,Reddy2024ed,Dong2023fqah,Zhou2023,DongJ2023,Guo2023,Kwan2023,Sheng2024,Song2024ahc,Tan2024,Soejima2024,DongZ2024,Shen2024,Kudo2024,Zeng2024ahc,Crepel2024,Yu2024,Huang2024FQAH}. Here, we investigate the interplay of electronic crystals with integer and fractional Chern insulators in a moir\'e lattice of rhomobohedral pentalayer graphene (RPG) aligned with hexagonal boron nitride. At a doping of one electron per moir\'e unit cell, we see a correlated insulator with a Chern number that can be tuned between $C=0$ and $+1$ by an electric displacement field, accompanied by an array of other such insulators formed at fractional band fillings, $\nu$. Collectively, these states likely correspond to trivial and topological electronic crystals, some of which spontaneously break the discrete translational symmetry of the moir\'e lattice. Upon applying a modest magnetic field, a narrow region forms around $\nu=2/3$ in which transport measurements imply the emergence of a fractional Chern insulator, along with hints of weaker states at other fractional $\nu$. In the same sample, we also see a unique sequence of incipient Chern insulators arising over a broad range of incommensurate band filling near two holes per moir\'e unit cell. Our results establish moir\'e RPG as a fertile platform for studying the competition and potential intertwining of electronic crystallization and topological charge fractionalization. 
}

Moir\'e materials with flat bands have become ideal platforms for studying the interplay between strongly correlated and topological states of matter~\cite{Balents2020,Andrei2020,Mak2022,Nuckolls2024,Adak2024}. Early work in this field established the existence of a wide range of correlation-driven topological states within the magnetic subbands of the Hofstadter butterfly spectrum, including states that break the translational symmetry of the moir\'e lattice (called ``symmetry-broken Chern insulators'') and others that exhibit fractionalized charge excitations (fractional Chern insulators, FCI)~\cite{Dean2013,Spanton2018}. Recently, remarkable progress has been made in achieving analogous states that appear even in the absence of an external magnetic field in moir\'e lattices with flat bands and suitable quantum geometry. Prominent examples include the integer and fractional quantum anomalous Hall (IQAH and FQAH) states found in twisted bilayer MoTe$_2$~\cite{Cai2023,Zeng2023,Park2023,Xu2023fqah} and moir\'e lattices of rhombohedral multilayer graphene with aligned hBN~\cite{Chen2020,Lu2024,Xie2024FQAH}, as well as correlated Chern insulator states with associated translational symmetry breaking found in a variety of other twisted graphene moir\'e lattices~\cite{xie2021fractional,polshyn2022topological,Su2024}. 

\begin{figure*}[t]
\includegraphics[width=\textwidth]{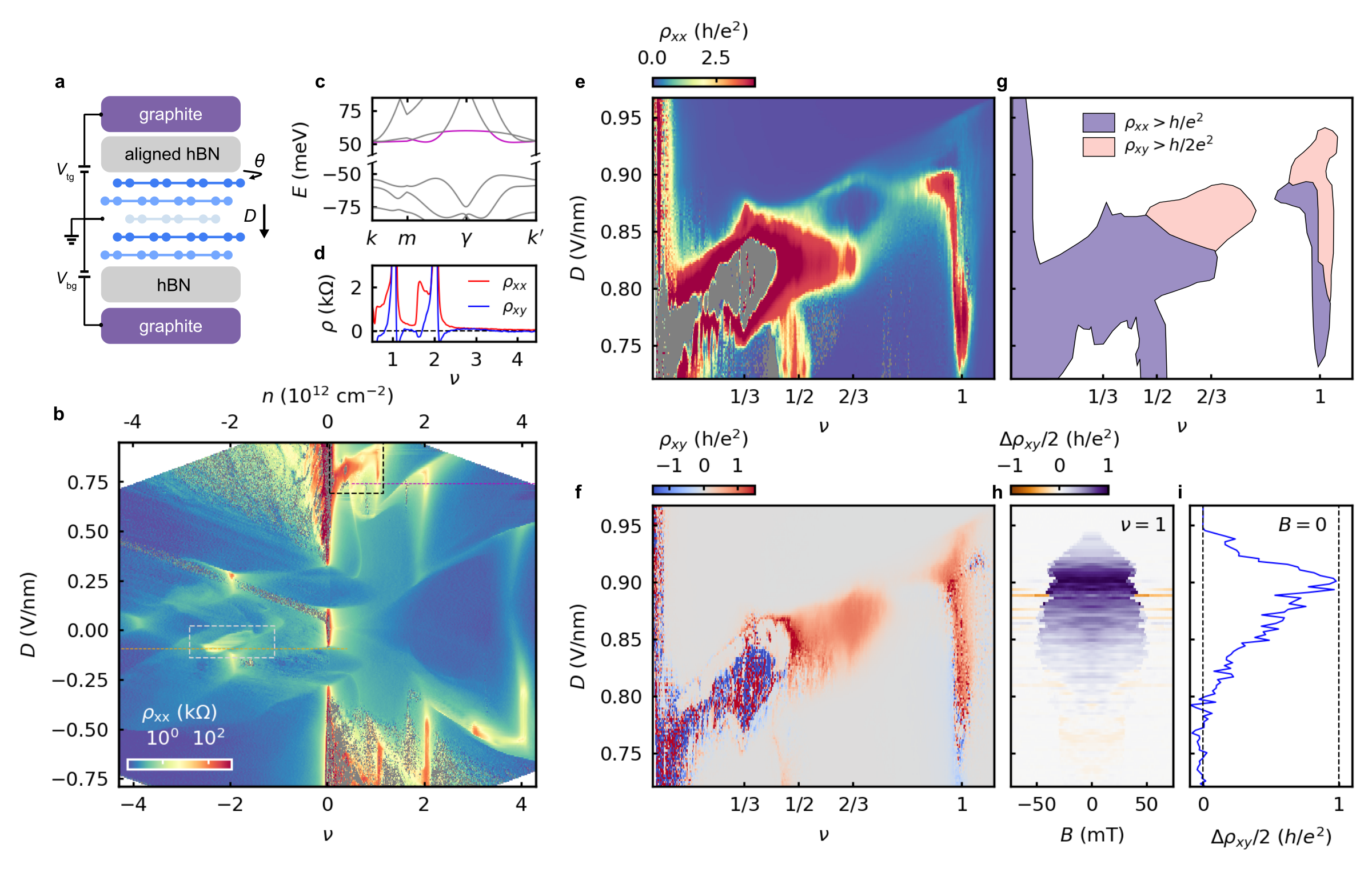} 
\caption{\textbf{Device transport characterization and correlated states at $\nu\leq1$.}
\textbf{a}, Schematic of the device. RPG is encapsulated between flakes of hBN with graphite top and bottom gates. The RPG is misaligned from the bottom hBN, and has a small twist angle of $\theta=0.90^{\circ}$ with the top hBN to form a moiré superlattice (see Methods for discussion).
\textbf{b}, Map of $\rho_{xx}$ taken at $B=0$ over a wide range of the $n-D$ parameter space. The data is acquired in multiple sub-measurements to mitigate measurement artifacts, as described in the Methods. The dashed black (gray) box indicates the region of data shown in \textbf{e-f} (Fig.~\ref{fig:4}a). The gray colored data in the color scale indicates that $\rho_{xx}$ was measured to be negative, indicating either a highly resistive state or poor equilibration of the contacts (see Methods).
\textbf{c}, Single-particle calculated band structure with $\delta=+150$~meV and a moir\'e period of $10.8$~nm. This sign of $\delta$ corresponds to $D>0$. The lowest moiré conduction band is colored in purple. 
\textbf{d}, Line traces of $\rho_{xx}$ and $\rho_{xy}$ acquired at $D=0.740$~V/nm (corresponding to the position of the purple dashed line in \textbf{b}) with $B=0.2$~T (not symmetrized).
\textbf{e}, Zoomed-in map of $\rho_{xx}$ symmetrized at $|B|=100$~mT from the region of the black dashed box in \textbf{b}.
\textbf{f}, Similar map of antisymmetrized $\rho_{xy}$.
\textbf{g}, Schematic indicating transport features seen in \textbf{e-f}. Regions shaded in pink satisfy the condition $\rho_{xy}>h/2e^2$. Regions shaded in purple satisfy the condition $\rho_{xx}>h/e^2$. Regions satisfying both conditions are also shaded in purple (since the large value of $\rho_{xy}$ is an artifact corresponding to mixing with $\rho_{xx}$). Regions with negative $\rho_{xx}$ are shaded in purple, as they correspond to measurement artifacts in very insulating states (see Methods). 
\textbf{h}, Map of $\Delta \rho_{xy}/2=\left(\rho_{xy}^{\uparrow}-\rho_{xy}^{\downarrow}\right)/2$ at $\nu=1$, where the arrows indicate the direction $B$ is swept. 
\textbf{i}, A line cut of $\Delta \rho_{xy}/2$ from \textbf{h} at $B=0$.
}
\label{fig:1}
\end{figure*}

Forming Chern insulators in moir\'e lattices requires spontaneous valley polarization, resulting in broken time-reversal symmetry~\cite{Nuckolls2024,Adak2024}. Many-body gaps can be opened at integer fillings of the moir\'e flat bands, resulting in the emergence of Chern insulators and the IQAH effect when the filled bands have nonzero total Chern number. Topological gapped states can also form at fractional fillings of the \moire bands, but require the assistance of additional correlation-driven mechanisms. FQAH states are a notable example, forming anyonic quasiparticles but otherwise not needing to break any of the remaining symmetries of the system. Alternatively, electrons can spontaneously break translational symmetry to form topological electronic crystal (TEC) states. Examples include the anomalous Hall crystal (AHC) recently considered for moir\'e and non-moir\'e rhombohedral pentalayer graphene (RPG)~\cite{Dong2023fqah,Zhou2023,DongJ2023,Kwan2023,Tan2024,Soejima2024,DongZ2024,Shen2024,Zeng2024ahc,Kudo2024,Crepel2024}, and the generalized AHC seen in twisted bilayer-trilayer graphene~\cite{Su2024}. Understanding the interplay between TEC states and FQAH states stands as a critical open challenge for the field.

Here, we study an array of correlation-driven topological states arising in a RPG/hBN moir\'e lattice with a period of $10.8$~nm. Recent pioneering work on this system revealed the emergence of FQAH states at several Jain-sequence fractions for $\nu<1$, all arising in a device with a slightly larger moir\'e period of $11.5$~nm~\cite{Lu2024}. In contrast, our device instead exhibits a sequence of trivial insulators and IQAH states, with Chern numbers of $C=0$ and $+1$, originating from both integer and fractional fillings of the electron-doped moir\'e conduction band. These states most likely correspond to trivial ($C=0$) and topological ($C=+1$) electronic crystals, separated by a phase transition controlled by doping, $n$, electric displacement field, $D$, and out-of-plane magnetic field, $B$. They are most robust near commensurate band fillings ($\nu=4n/n_s =1/3$, $2/3$, and $1$, where $n_s$ is the \moire superlattice density, see Methods), but nevertheless extend over a wide range of moir\'e band filling, $0 < \nu \lessapprox 1$. Strikingly, there are clear signatures of an incipient FCI state that emerges in a magnetic field of around 2 T and originates from $\nu=2/3$, along with hints of other Jain-sequence states. In a different part of the gate-tuned phase diagram, a set of unique topological states form in a distinct valley-polarized pocket near a doping of two holes per moir\'e unit cell ($\nu\approx-2$). There, we see an incipient Chern insulator emerging from an incommensurate filling of $\nu\approx-2.45$, as well as an unusual sequence of at least $10$ weaker topological states arising as the hole doping is reduced. Collectively, our observations pose several intriguing new puzzles for future studies.

\medskip\noindent\textbf{Trivial and topological electronic crystals}

Figure~\ref{fig:1}a shows a schematic of our device, in which RPG is nearly aligned to one hBN dielectric but misaligned from the other. The top and bottom graphite gates enable independent control over $n$ and $D$, the latter of which we define to be positive when electrons are pushed away from the \moire interface. Figure~\ref{fig:1}b shows a map of the longitudinal resistance, $\rho_{xx}$, of the device taken over a wide range of $n$ and $D$. We extract the density needed to fully fill the lowest moir\'e bands ($\nu=\pm4$) to be $n_s=3.98\times10^{12}$~cm$^{-2}$, corresponding to a moir\'e period of $10.8$~nm and a twist angle between the RPG and hBN of $\theta=0.90^{\circ}$ (assuming a 1.7\% lattice mismatch between graphene and hBN). Overall, the salient features we see are consistent with those reported in a prior study of moir\'e RPG~\cite{Lu2024}, including a series of insulating states at the charge neutrality point ($\nu=0$) and resistive states at various integer values of $\nu$ corresponding to correlated or single-particle band insulators. In the Methods section, we provide additional supporting details for all of the descriptions and analysis above (see also Extended Data Figs.~\ref{efig:device_fab}-\ref{efig:basic_fans}).

Figure~\ref{fig:1}c shows a continuum model band structure calculation for an interlayer potential difference of $\delta=+150$~meV (see Methods). The calculation predicts that the lowest \moire conduction band (colored in purple) is gapped from the highest \moire valence band, but overlaps the second conduction band. Both of these features are consistent with the experiment, in which there is an insulating state at $\nu=0$ but a metallic state at $\nu=4$ for large positive $D$ (see Fig.~\ref{fig:1}d for a representative measurement at $D=0.740$~V/nm, corresponding to the purple dashed line in Fig.~\ref{fig:1}b). Figures~\ref{fig:1}e-f show zoomed-in maps of the field-symmetrized longitudinal ($\rho_{xx}$) and anti-symmetrized Hall ($\rho_{xy}$) resistances in the high-$D$ region between $\nu=0$ and $\nu\approx1$, outlined by the black dashed square in Fig.~\ref{fig:1}b. The key features seen in these maps are summarized in Fig.~\ref{fig:1}g, where regions of $\rho_{xy}>h/2e^2$ are colored in pink and regions of $\rho_{xx}>h/e^2$ (or regions that are artificially negative owing to their extremely high resistance) are colored in purple. 

\begin{figure*}[t]
\includegraphics[width=\textwidth]{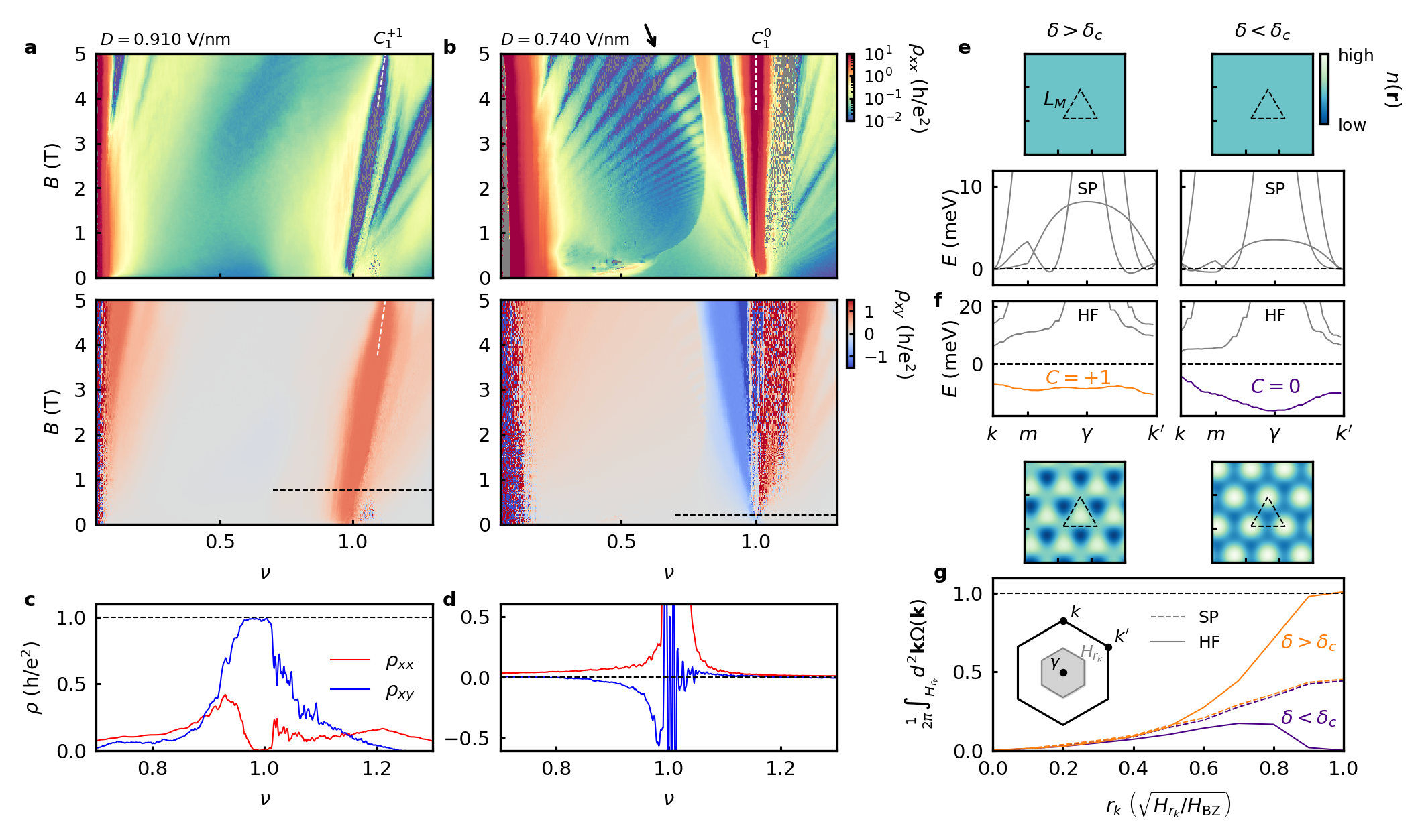} 
\caption{\textbf{Topological and trivial electronic crystals at $\nu=1$.}
\textbf{a}, Landau fan diagram of $\rho_{xx}$ (top) and $\rho_{xy}$ (bottom) taken at $D=0.910$~V/nm. The white dashed line shows the expected evolution of a $C=+1$ state originating from $\nu=1$ based on the Streda formula.
\textbf{b}, Similar Landau fan taken at $D=0.740$~V/nm. The white dashed line shows the expected evolution of a $C=0$ state originating from $\nu=1$. The black arrow denotes the trajectory of a quantum Hall state with filling factor of $-3$ originating from $\nu=1$. The speckled features projecting vertically near $\nu=1$ are artifacts due to a large contact resistance.
\textbf{c}, Line traces of $\rho_{xx}$ and $\rho_{xy}$ taken at $B=0.75$~T from the $D=0.910$~V/nm Landau fan, as indicated by the black dashed line in \textbf{a}.
\textbf{d}, Similar line cuts taken at $B=0.20$~T from the $D=0.740$~V/nm Landau fan, as indicated by the black dashed line in \textbf{b}.
\textbf{e}, Single-particle (SP) calculations of the spatial distribution of carrier density, $n(\mathbf{r})$, at a filling of $\nu = 1$. The left (right) panel is calculated with $\delta=150$~meV ($120$~meV). Both correspond to metallic states. The associated band structure calculations are shown below each plot. The calculation is performed for a moir\'e period of $L_M=11.1$~nm with the moir\'e potential strength artificially set to zero (see Methods).
\textbf{f}, Similar calculations performed with the Hartree--Fock (HF) method. Both correspond to insulating states, with the filled band having $C=+1$ ($0$) for $\delta=150$~meV ($120$~meV), shown in orange (purple). Dashed lines at zero energy in \textbf{e-f} denote the Fermi energy at $\nu=1$. The real space densities in \textbf{e-f} share the same color scale. The dashed black triangles indicate the \moire unit cell. 
\textbf{g}, Berry curvature integrated from the center to the edge of the moir\'e Brillouin zone (BZ) for both values of $\delta$ considered in \textbf{e-f}. The dashed (solid) curves show the single-particle (Hartree--Fock) calculations. The inset schematic shows the \moire BZ in black, with high symmetry points labeled. The gray area depicts the area of integration, which scales with $r_k$, such that $H_{r_k}=r^2_kH_{\mathrm{BZ}}$, where $H_{\mathrm{BZ}}$ is the area of the full \moire BZ.
}
\label{fig:2}
\end{figure*}

The behavior of $\rho_{xx}$ and $\rho_{xy}$ change substantially with both $n$ and $D$. There is a stripe-like region cutting diagonally across the center of the maps in Figs.~\ref{fig:1}e-f, in which one or both of $\rho_{xx}$ and $\rho_{xy}$ are very large. The system is a trivial insulator for $\nu\lessapprox 1/2-2/3$ (depending on the precise value of $D$), previously attributed to the formation of a Wigner crystal with period larger than the original moir\'e lattice~\cite{Lu2024}. There is also a pocket centered at $\nu=2/3$ featuring large $\rho_{xy}$ and a deep suppression of $\rho_{xx}$. This behavior instead indicates the formation of a topological gapped state. At $\nu=1$, there is an extended vertical transport feature that exhibits a large anomalous Hall effect (AHE) for $0.82$~V/nm~$\lessapprox D \lessapprox 0.95$~V/nm but diverging $\rho_{xx}$ with no AHE for $0.73$~V/nm~$\lessapprox D \lessapprox 0.82$~V/nm (see Figs.~\ref{fig:1}h-i).

To more clearly probe the nature of these states, we plot Landau fan diagrams of $\rho_{xx}$ and $\rho_{xy}$ for $D=0.910$~V/nm and $D=0.740$~V/nm (Figs.~\ref{fig:2}a and b, respectively; see Extended Data Fig.~\ref{efig:unsymmetrized_fans} for additional Landau fans from multiple voltage probes). In the former, we see a correlated Chern insulator emerging from $B=0$ at $\nu=1$, along with additional associated quantum Hall states emerging to its right at higher field. The line cut acquired at $B=0.75$~T in Fig.~\ref{fig:2}c confirms that the correlated Chern insulator state exhibits the anticipated values of $\rho_{xx}\approx 0$ and $\rho_{xy}\approx h/e^2$, where $h$ is Planck's constant and $e$ is the charge of the electron. Comparison to the Streda formula~\cite{streda1982quantised, streda1982theory}, $\left(\frac{\partial n}{\partial B}\right)_{\mu}=C\frac{h}{e}$, further confirms that the Chern number of this state is $C=+1$, as indicated by the white dashed line shown at the top of the Landau fans. We label gapped states by their Chern number and band filling upon extrapolation to $B=0$ following the convention $(C=a, \nu=b) \equiv C^{a}_{b}$, such that this state is $C^{+1}_{1}$. The insulating behavior onsets below a temperature of $T \approx 3$~K with a gap $\Delta_{\nu=1} = 96$~$\mu$eV ($149$~$\mu$eV) at $D=0.909$~V/nm, determined from the temperature dependence of $\rho_{xx}$ ($\rho_{xy}$) (see Extended Data Fig.~\ref{efig:AHC_temp_dependence}a).

The Landau fan taken at $D=0.740$~V/nm exhibits markedly different behavior. At $\nu=1$, $\rho_{xx}$ far exceeds $h/e^2$ and $\rho_{xy}$ exhibits diverging behavior with an abrupt sign reversal across integer filling (see line cuts in Fig.~\ref{fig:2}d taken at $B=0.20$~T). The insulating state does not disperse with $B$, and associated quantum Hall states emerge roughly symmetrically to its left and right. Collectively, this behavior is consistent with a topologically trivial state, $C^{0}_{1}$. The value of $D\approx 0.82$~V/nm where the AHE at $\nu=1$ vanishes most likely corresponds to a phase transition between two topologically distinct states with associated Chern numbers of $C=0$ and $+1$. Such a phase transition is typically first-order and requires a gap closure; we do not see evidence for either in our device, potentially due to disorder.

Many recent theory works~\cite{Dong2023fqah,Zhou2023,DongJ2023,Tan2024,Soejima2024,DongZ2024,Shen2024,Zeng2024ahc,Kudo2024,Crepel2024,Yu2024,Huang2024FQAH} have considered the nature of the Chern insulator previously reported at $\nu=1$~\cite{Lu2024}. In many other moir\'e systems, a correlated gap opens at $\nu=1$ when the four-fold isospin degeneracy is lifted by interactions \cite{Nuckolls2024,Adak2024}. But in RPG, the second moir\'e conduction band overlaps the first; as a result, this mechanism by itself would not be expected to open a gap. The AHC has thus been proposed as a new mechanism for generating a Chern insulator, in which the additional formation of an electronic crystalline order via a large spatial redistribution of the charge density opens the topological gap~\cite{Dong2023fqah,Zhou2023,DongJ2023,Kwan2023,Tan2024,Soejima2024,DongZ2024,Shen2024,Zeng2024ahc,Kudo2024,Crepel2024}. A definitive understanding of the gap at $\nu=1$ remains elusive, however, since the putative electronic crystal is commensurate with the moir\'e lattice and thus challenging to distinguish unambiguously from conventional moir\'e Chern insulators~\cite{Parameswaran2024}. Nevertheless, the basic TEC framework is consistent with the $C=1$ state we observe at $\nu=1$. The trivial $C=0$ insulator at smaller $D$ can be formed by lifting the same isospin degeneracies and crystallizing the electrons as for the AHC, but with the filled states instead having a total Chern number of zero. 

\begin{figure*}[t]
\includegraphics[width=\textwidth]{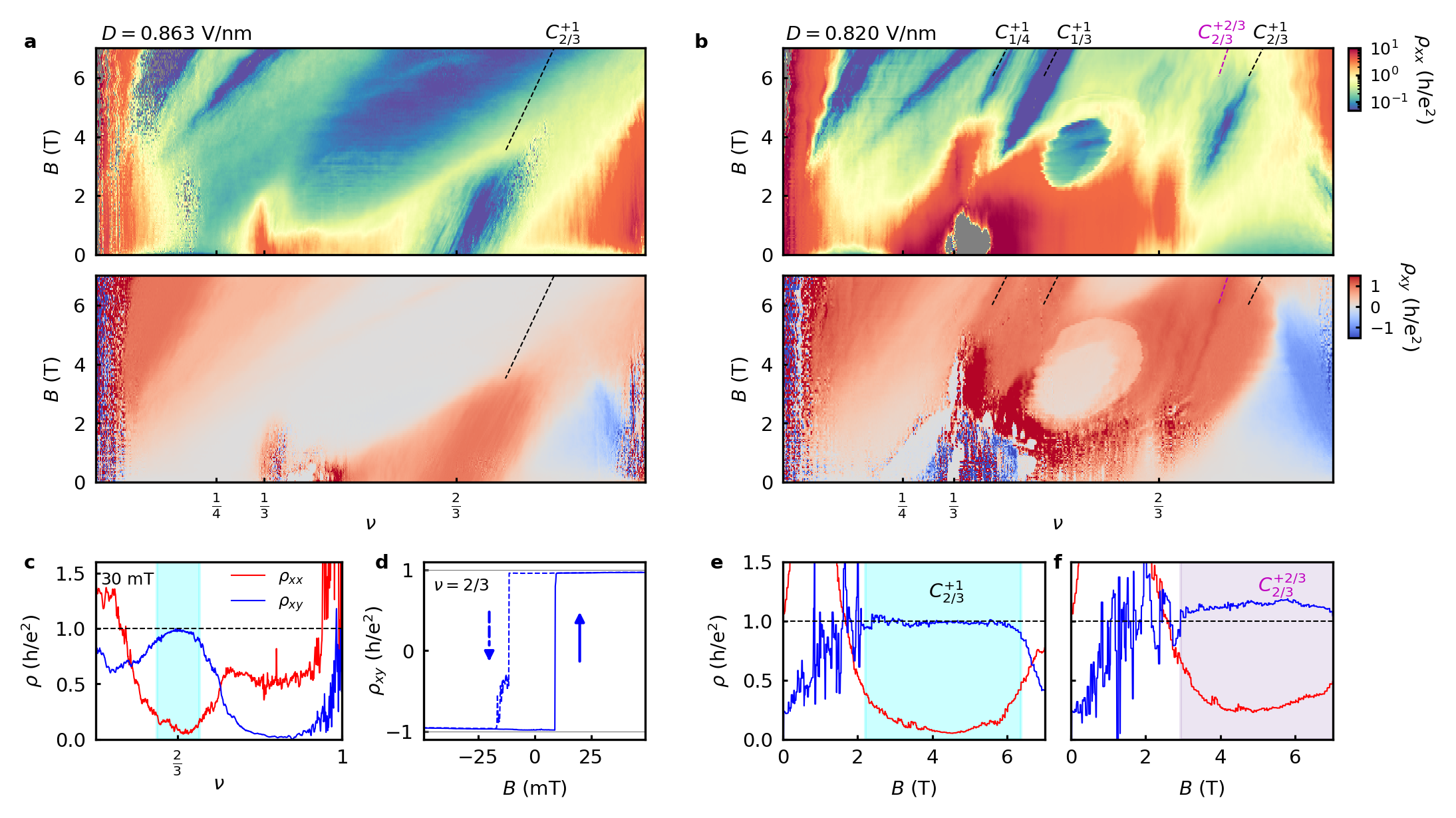} 
\caption{\textbf{Competing TEC and FCI states at fractional $\nu$.}
\textbf{a}, Landau fan diagram of $\rho_{xx}$ (top) and $\rho_{xy}$ (bottom) taken at $D=0.863$~V/nm. The black dashed line shows the expected evolution of a $C=+1$ state originating from $\nu=2/3$ based on the Streda formula.
\textbf{b}, Similar Landau fan taken at $D=0.820$~V/nm. The black dashed lines show the expected evolution of $C=+1$ states originating from $\nu=1/4$, $1/3$, and $2/3$. The purple dashed line shows the same for a $C=+2/3$ state originating from $\nu=2/3$.
\textbf{c}, Line traces of $\rho_{xx}$ and $\rho_{xy}$ taken at $B=30$~mT from the $D=0.863$~V/nm Landau fan in \textbf{a}. The blue shaded region corresponds to the contiguous range of $\nu$ for which $\rho_{xy}>0.9h/e^2$.
\textbf{d}, Measurement of $\rho_{xy}$ at $\nu=2/3$ acquired as $B$ is swept back and forth across zero. Arrows denote the sweep direction of the magnetic field.
\textbf{e}, Line traces of $\rho_{xx}$ and $\rho_{xy}$ taken along the trajectory indicated by the black dashed line associated with the $C^{+1}_{2/3}$ state in \textbf{b}. The blue shaded region corresponds to the contiguous region of $|\rho_{xy}|>0.9 h/e^2$ (excluding $B<2$~T which is dominated by the trivial insulating phase).
\textbf{f}, Similar line traces taken along the trajectory indicated by the purple dashed line associated with the $C^{+2/3}_{2/3}$ state in \textbf{b}. The purple shaded region corresponds to contiguous range of $\nu$ for which $\rho_{xy}>h/e^2$ (excluding $B<3$~T).
}
\label{fig:3}
\end{figure*}

To better understand the possible nature of the gapped states at $\nu=1$, we compare our results to Hartree--Fock (HF) calculations performed with the moir\'e potential artificially set to zero (see Methods for details). Figures~\ref{fig:2}e-f show the calculated band structure at $\nu=1$ both before and after HF for two different values of $\delta$. In both cases, the state is gapless at the single-particle level but gapped as a result of interactions. The isolated band has a Chern number of $C=0$ for $\delta$ less than a critical value of $\delta_c\approx 140$~meV, but $C=+1$ for $\delta > \delta_c$. This is consistent with our observed phase transition between a $C=0$ and $+1$ state as $D$ is increased. At the non-interacting level, the integrated Berry curvature up to the Fermi level at $\nu=1$ can be any arbitrary value since the band is not isolated. Figure~\ref{fig:2}g shows that this value is $\frac{1}{2\pi}\int_{\mathrm{BZ}}d^2\mathbf{k}\Omega(\mathbf{k}) \approx 0.5$ in the calculations for both $\delta=120$~meV and $150$~meV. When interactions open a gap, however, the filled states below the Fermi level must have Berry curvature that integrates to a quantized value, equal to $C$. In this context, the gap opening simultaneously necessitates an interaction-driven modification of the Berry curvature in order to satisfy the quantization condition. Small changes in the quantum geometry of the single-particle bands with $\delta$ can thus lead to an abrupt phase transition between otherwise similar states having $C=0$ (e.g., a generalized Wigner crystal) and $C=1$ (e.g., a generalized anomalous Hall crystal, where ``generalized'' indicates the role of the moir\'e potential in seeding the crystal formation~\cite{Regan2020,Su2024}). Combined with the apparent band overlap indicated by the metallic behavior at $\nu=4$ (Fig.~\ref{fig:1}d), this framework provides a compelling explanation for our observations at $\nu=1$. Nevertheless, future scanning probe microscopy studies will be needed to directly confirm this interpretation. 

\medskip\noindent\textbf{Interplay of electronic crystals and FCI}

Turning now to the correlated states at fractional band filling, we examine Landau fans acquired at intermediate values of $D$ where noteworthy transport features are most pronounced (Figs.~\ref{fig:3}a-b for $D=0.863$~V/nm and $0.820$~V/nm, respectively). Figure~\ref{fig:3}a cuts through the largest values of $R_{xy}$ in the $n-D$ map, and exhibits a broad region of suppressed $\rho_{xx}$ with concomitant $\rho_{xy}\approx h/e^2$ surrounding $\nu=2/3$ (see line cuts in Fig.~\ref{fig:3}c). At $\nu=2/3$, $\rho_{xy}$ switches between $\approx\pm h/e^2$ in a single hysteresis loop with a small coercive field of $B\approx 10$~mT (Fig.~\ref{fig:3}d). The slope of this state in the Landau fan is consistent with a $C=+1$ Chern insulator based on the Streda formula (i.e., $C^{+1}_{2/3}$), as indicated by the black dashed line in Fig.~\ref{fig:3}a. These features are plainly incompatible with the $C=2/3$ FQAH state at $\nu=2/3$ reported previously~\cite{Lu2024}, since our observed state has neither the appropriate Streda slope in the Landau fan nor the appropriate quantization of $\rho_{xy}$ for a $C=2/3$ state (that is, we find $\rho_{xy}=h/e^2$ rather than $3h/2e^2$). Instead, the $C^{+1}_{2/3}$ state most naturally corresponds to a TEC that spontaneously enlarges the unit cell area (see Methods for a discussion of two possible topological orders of this crystalline state). Notably, $\rho_{xy}$ remains large over a relatively wide range of doping, potentially indicating that the crystalline order persists in some form even upon doping away from $\nu=2/3$. The insulating behavior onsets below $T \approx 2$~K with a gap $\Delta_{\nu=2/3} = 54$~$\mu$eV ($44$~$\mu$eV) at $D=0.870$~V/nm, determined from the temperature dependence of $\rho_{xx}$ ($\rho_{xy}$) (see Extended Data Fig.~\ref{efig:AHC_temp_dependence}b).  

Similar behavior is seen in the Landau fan at $D=0.820$~V/nm, although a magnetic field is required to induce the topological phase transition from trivial to Chern insulator (see the line cut in Fig.~\ref{fig:3}e). Two additional translational symmetry--broken states with $C=+1$ emerge at even higher magnetic field, projecting to $\nu=1/4$ and $\nu=1/3$ at $B=0$ (i.e., $C^{+1}_{1/4}$ and $C^{+1}_{1/3}$). There is also an oval-shaped feature near $\nu\approx0.6$ centered at $B\approx3.5$~T, in which both $\rho_{xx}$ and $\rho_{xy}$ are abruptly suppressed and there are instead quantum Hall states projecting to the CNP at $\nu=0$. This region is separated from the surrounding area of the Landau fan by a first-order phase transition and has the same origin as the sharp curved feature in the Landau fan in Fig.~\ref{fig:2}b. This phase transition potentially reflects a collapse of the electronic crystalline order (see Methods and Extended Data Fig.~\ref{efig:fan_hysteresis} for additional details).

Remarkably, the Landau fan in Fig.~\ref{fig:3}b also contains an extremely narrow feature in which the antisymmetrized $\rho_{xy}$ exceeds $h/e^2$ (see the line cut in Fig.~\ref{fig:3}f). The purple dashed line near the top of the Landau fan denotes the position and trajectory of this state, which projects precisely to $\nu=2/3$ at $B=0$ and has a Streda slope consistent with a $C=+2/3$ state (i.e., $C^{+2/3}_{2/3}$). These features are consistent with an incipient $\nu=2/3$ FCI state emerging with $B$. Although the state is not fully developed, with large residual $\rho_{xx}$ and non-quantized $\rho_{xy}$, the observation of $\rho_{xy}>h/e^2$ with a Streda slope implying $C=+2/3$ has no simple explanation besides a field-induced FCI state. This putative FCI appears more clearly in similar Landau fans taken at 300~mK, as well as in corresponding maps of the Hall angle, $\theta_H=\arctan(\rho_{xy}/\rho_{xx})$ (see Extended Data Figs.~\ref{efig:fan_with_Hall_angles}-\ref{efig:other_fan_with_Hall_angles}). The Landau fan shown in Extended Data Fig.~\ref{efig:other_fan_with_Hall_angles} additionally carries hints of several other incipient Jain-sequence FCI states originating from $\nu=1/3$, $2/5$, $3/7$, $4/7$, and $3/5$, all with $R_{xy}>h/e^2$ and a relatively large $\theta_H$. Further hints of intermixed FCI and TEC states can be seen in transport maps of $\rho_{xx}$ and $\rho_{xy}$ taken at $|B|=2$~T (Extended Data Fig.~\ref{efig:2T_map}), which exhibit broad regions of $\rho_{xy}>h/e^2$ with interspersed pockets of $\rho_{xy}\approx h/e^2$.

\begin{figure*}[t]
\includegraphics[width=\textwidth]{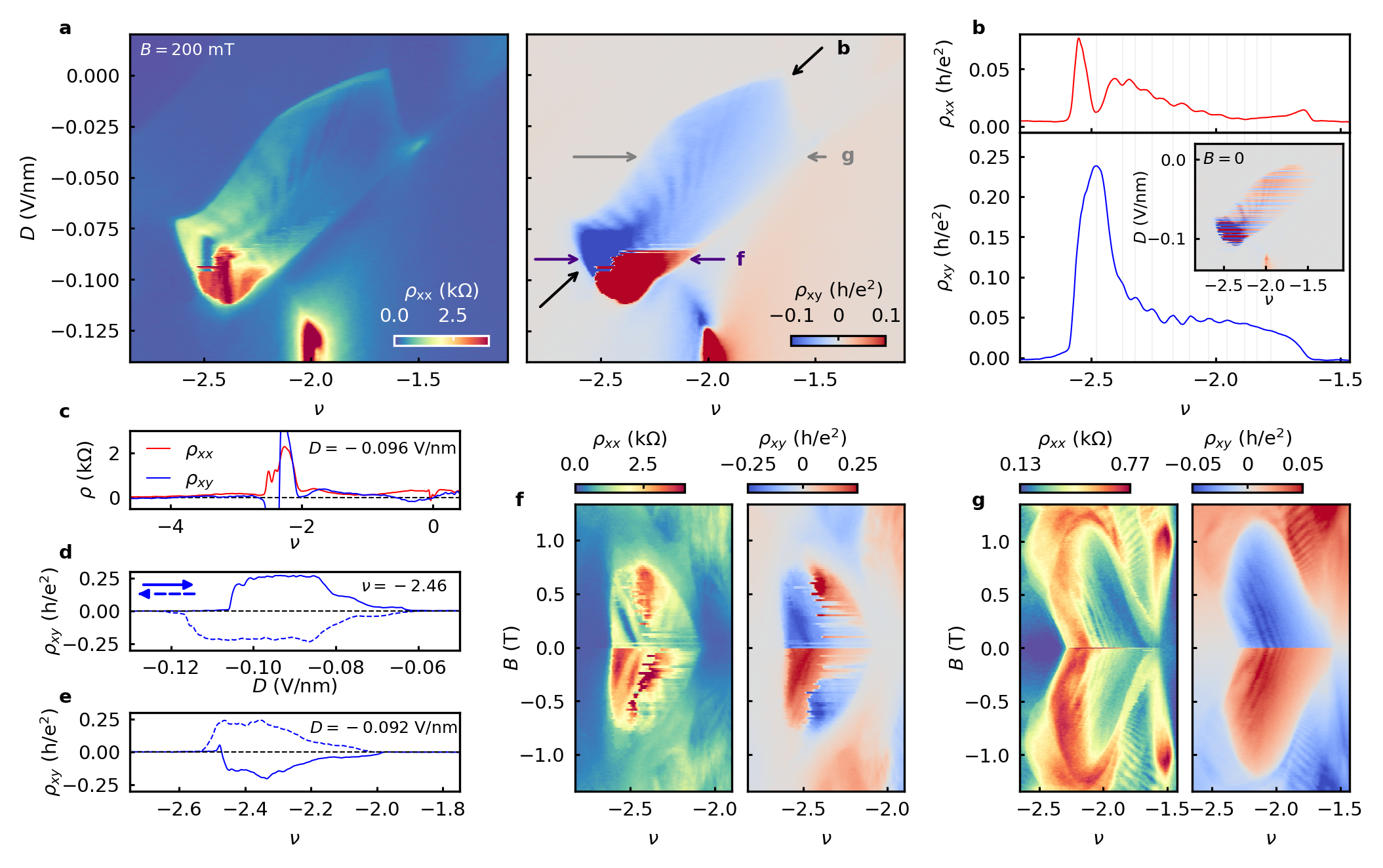} 
\caption{\textbf{Unusual topological states in a valley-polarized pocket near $\nu=-2$.}
\textbf{a}, Zoomed-in maps of $\rho_{xx}$ (left) and $\rho_{xy}$ (right) taken at $B=200$~mT (not symmetrized) from the region of the gray dashed box in Fig.~\ref{fig:1}b.
\textbf{b}, Line traces of $\rho_{xx}$ and $\rho_{xy}$ taken along a diagonal trajectory indicated by the black arrows in \textbf{a}. Vertical gray lines denote values of $\nu$ with concomitant $\rho_{xx}$ minima and $\rho_{xy}$ maxima. We plot the absolute value of $\rho_{xy}$ for clarity. The inset shows the same $\rho_{xy}$ map as in \textbf{a} but measured at $B=0$.
\textbf{c}, Line traces of $\rho_{xx}$ and $\rho_{xy}$ acquired at $B=0.2~\rm{T}$ and $D=-0.096$~V/nm (corresponding to the position of the yellow dashed line in Fig.~\ref{fig:1}b).
\textbf{d}, $\rho_{xy}$ trace measured at fixed $\nu=-2.46$ as $D$ is swept back and forth with $B=0$.
\textbf{e}, $\rho_{xy}$ trace measured at fixed $D=-0.092$~V/nm as $\nu$ is swept back and forth with $B=50$~mT.
\textbf{f}, Landau fan diagram of $\rho_{xx}$ (left) and $\rho_{xy}$ (right) taken at $D=-0.090$~V/nm, as indicated by the purple arrows in \textbf{a}.
\textbf{g}, Similar Landau fan diagram taken at at $D=-0.040$~V/nm, as indicated by the gray arrows in \textbf{a}.
}
\label{fig:4}
\end{figure*}

\medskip\noindent\textbf{Topological states at fractional hole doping}

In addition to the valley-polarized states considered so far for electron-type doping, the device exhibits a second distinct pocket of valley polarization featuring the AHE near a doping of two holes per moir\'e unit cell ($\nu\approx-2$). This pocket arises for $D<0$ (gray dashed box in Fig.~\ref{fig:1}b), such that holes are pushed to the same hBN interface as for the electron-doped region considered earlier. The transport properties of this region are shown in maps taken at $B=200$~mT (Fig.~\ref{fig:4}a). The valley-polarized pocket resides close to a trivial ($C=0$) correlated insulator at $\nu=-2$ and slightly larger $|D|$, although the latter appears to be a completely distinct and unrelated state (see Methods). Overall, the behavior of this valley-polarized pocket is highly atypical. First, it has no particularly noteworthy features at the exact integer filling of $\nu=-2$. Instead, the predominant feature is a deep suppression of $\rho_{xx}$ and a large enhancement of $\rho_{xy}$ to $\approx h/4e^2$ at an incommensurate fractional filling of $\nu\approx-2.45$ (Fig.~\ref{fig:4}b). Second, a sequence of at least 10 additional oscillations in both $\rho_{xx}$ and $\rho_{xy}$ emerge upon reducing the hole doping. These can be seen in the maps in Fig.~\ref{fig:4}a, as well as in the line cut shown in Fig.~\ref{fig:4}b taken along the diagonal trajectory indicated by the black arrows in Fig.~\ref{fig:4}a. Third, the valley-polarized pocket appears to arise even though the highest moir\'e valence band is not isolated from other bands. Figure~\ref{fig:4}c shows measurements of $\rho_{xx}$ and $\rho_{xy}$ across a wide range of hole doping at $D=-0.096$~V/nm (see also Fig.~\ref{fig:1}b). The measurement reveals that both $\nu=0$ and $-4$ remain metallic at a value of $D$ for which the valley-polarized pocket is prominent. Combined with additional measurements revealing a complex sequence of quantum oscillations near $\nu=-2$ (Extended Data Fig.~\ref{efig:mulitband_transport}), we conclude that the highest moir\'e valence band may coexist with other moir\'e bands when the valley-polarized pocket is formed.

The oscillation features in the valley-polarized pocket persist to $B=0$ (inset of Fig.~\ref{fig:4}b), although they are less clear owing to random switching of the background orbital magnetic state \cite{Polshyn2020, Grover2022}. The magnetic switching is a ubiquitous feature throughout the valley-polarized pocket, with hysteretic sign changes in $\rho_{xy}$ driven by separately sweeping either $n$ or $D$ (Fig.~\ref{fig:4}d-e and Extended Data Fig.~\ref{efig:orbital_magnet}). Each of the oscillations corresponds to a local minimum of $\rho_{xx}$ concomitant with a maximum of $\rho_{xy}$, hinting at a topological nature of these states. Landau fans reveal that these oscillatory features drift to different $\nu$ as $B$ is raised (Figs.~\ref{fig:4}f-g), providing further evidence of their nontrivial topology. However, the properties of these states are highly unusual. The slope of each state is very large, often implying a Chern number as large as $13$ based on a naive application of the Streda formula. These trajectories are also not perfectly linear, indicating a breakdown of the applicability of the Streda formula. Additionally, whereas the primary state remains pinned to $\nu \approx -2.45$ upon varying $D$, all of the other states bear no fixed relationship between $\nu$ and $D$. Instead, they exhibit non-uniform spacing and drift as a function of $D$ with no easily quantifiable pattern.

Although the nature of these states remains both mysterious and unprecedented in comparison to all previously known Chern insulators, we speculate on a possible explanation of their origin. In a simple model with band overlap, itinerant carriers from other bands can remain at the Fermi surface even when a valley-polarization gap opens in the highest moir\'e valence band. This multi-band behavior could explain the non-quantized $\rho_{xy}$ and large residual $\rho_{xx}$, since the topological protection of the gapped states is spoiled by itinerant carriers in other bands. Such a scenario also necessitates modifying the Streda formula, as described in the Methods, which could explain the nonlinear trajectories of some of these states in the Landau fans. Finally, it implies that the nominal value of $\nu$ in fact reflects the total filling of all the relevant subbands, rather than the highest moir\'e valence band alone. In such a picture, the primary state seen at $\nu\approx-2.45$ may actually correspond to a filling of precisely two holes per moir\'e unit cell in the highest moir\'e valence band, with the remaining holes residing in the overlapping bands. The sequence of additional states arising at reduced hole doping may be a periodic recurrence of the primary Chern state, arising as interactions induce non-uniform charge transfer between the bands~\cite{kolavr2023electrostatic}.  

\medskip\noindent\textbf{Discussion and outlook}

Overall, our results establish moir\'e RPG as a promising platform for investigating the interplay between TEC and FCI states. Understanding their relationship stands as a critical open challenge for the field, especially as it relates to determining the ultimate ground state ordering of the system. Our observations further motivate the tantalizing possibility of finding topological states with coexisting charge fractionalization and electronic crystallization. Understanding the dependence of the TEC and FCI states on the moir\'e period, as well as studying the impact of Coulomb screening~\cite{Liu2021screening}, may help to further elucidate their relationship. 

In our measurements, we find likely TEC states at both fractional and integer fillings forming above $D \approx 0.82-0.87$~V/nm (depending on $\nu$). The most likely ground state for smaller $D$ is a generalized Wigner crystal with $C=0$, but this undergoes a phase transition to various topological phases above $B \approx 1-3$~T (depending on $D$, see Figs.~\ref{fig:1}e-g, Figs.~\ref{fig:3}a-b, and Extended Data Figs.~\ref{efig:fan_with_Hall_angles}-\ref{efig:2T_map}). These topological phases arise as intermixed integer and incipient fractional Chern insulators, pointing to a close and possibly intertwined relationship between the two. Notably, the putative generalized Wigner crystal extends over the largest range of $D$ very near $\nu=1/3$ (Fig.~\ref{fig:1}e). Similarly, both $C=0$ and $C=+1$ insulators appear most robustly very near $\nu=2/3$ and $\nu=1$ (Figs.~\ref{fig:2}a-b and Fig.~\ref{fig:3}a), collectively indicating that the electronic crystals are most stable at commensurate fillings of the moir\'e lattice.

Looking forward, we note that the competing trivial and topological electronic crystals at a given $\nu$ are likely to feature different spatial structures owing to the delocalized nature of the topological states in real space (see Fig.~\ref{fig:3}f). Scanning probe studies will be crucial for directly imaging the electronic crystals and better understanding the nature of their doping-dependent translational symmetry breaking. The other pocket of valley-polarization near two holes per moir\'e unit cell also holds many intriguing mysteries, featuring signatures of unusual topological behaviors not previously seen in other systems. Twist-angle--dependent studies will be valuable, as this pocket may share the same origin as the multiferroic `bubble' region previously studied in RPG without a moir\'e~\cite{Han2023}. From this perspective, the addition of the moir\'e potential may primarily serve to trigger the formation of a peculiar sequence of topological states within the valley-polarized pocket. Whether electronic crystallization is also needed to form these states stands as an open question for future study.

\section*{Methods}

\textbf{Identification of rhombohedral domains.} We identified five-layer thick flakes of exfoliated graphene using an optical microscope. We then mapped the stacking domains (e.g., Bernal and rhombohedral, along with other less energetically favorable configurations) with amplitude-modulated Kelvin probe force microscopy (AM-KPFM) using a Bruker Icon atomic force microscope (AFM) with an SCM-PIT-V2 tip (see Extended Data Fig.~\ref{efig:device_fab}). We applied an ac voltage bias to the tip between $500-1500$~mV and reduced the tip-sample separation by $5-10$~nm compared to the average height of the tapping mode. These parameters were adjusted during measurement in order to optimize the signal contrast. The tapping mode topography and AM-KPFM measurements were interleaved, such that the topography was collected during one raster direction of the tip and the AM-KPFM potentiometry was performed during the reverse raster scan (with the tip held at a constant tip-sample separation based on the measured topography). Although the graphene flake was not electrically grounded, which should preclude the applicability of AM-KPFM, we were nevertheless able to distinguish very clear contrast in the measured signal for different stacking domains. We could further confirm that a given domain of interest was indeed rhombohedral by comparing a Raman spectrum taken with a $514$~nm laser to the Bernal region along with other published curves~\cite{Han2024}. The process is shown in more detail in Supplementary Information Fig.~S1. The $C=-5$ Chern insulator seen in the Landau fan in Extended Data Fig.~\ref{efig:basic_fans} provides additional confirmation of both the layer number and rhombohedral stacking configuration of the graphene flake~\cite{Han2024}.

\textbf{Device fabrication.} The RPG region of interest was first isolated using a resist-free local anodic oxidation nanolithography process \cite{Li2018}, substantially reducing the chance that it relaxed into the energetically-favored Bernal stacking configuration as it was picked up into the vdW stack. In order to further reduce the risk of relaxation, we separately prepared the bottom portion of the stack by picking up an hBN and graphite dielectric/back-gate and placing them onto an SiO$_2$ substrate. We then cleaned the surface of the hBN using contact-mode AFM (using a OTESPA-R3 AFM tip and a line spacing of $\approx 100$ nm). This pre-assembly procedure allowed us to deposit the top half of the stack --- graphite/hBN/RPG --- as the last step of the vdW heterostructure assembly process, mitigating the risk of relaxing the rhombohedral flake upon further straining the crystal while picking up additional vdW flakes~\cite{Zhou2021_abctrilayer}. For all steps of stacking, we used standard dry transfer techniques with a polycarbonate (PC) film on a polydimethyl sioloxane (PDMS) stamp. 

Straight edges of the RPG were aligned to the top and bottom hBN, such that there was a 50\% chance of either hBN creating a long-wavelength moir\'e potential. We note that this technique led to an unintentional ambiguity in determining which hBN interface is aligned, as both in principle could be. Since the transport features of our device closely resemble many of those seen in Ref.~\cite{Lu2024}, we assume that we most likely have the same polarization scheme (i.e., both electrons and holes biased to the RPG surface away from the aligned hBN moir\'e interface in their respective valley-polarized pockets) and infer that the top hBN creates the long-wavelength moir\'e pattern. We only observe signatures of a single long-wavelength moir\'e pattern in the transport data up to a doping of $n=11.59\times10^{12}$~cm$^{-2}$ (Supplementary Information Fig.~S2), setting a lower bound on the twist angle of the misaligned hBN of $2.03^{\circ}$ and an upper bound on the shorter moir\'e period of $6.31$~nm.

After stacking, standard device fabrication procedures were employed to create the dual-gated Hall bar device (i.e., reactive ion etching and evaporation of 7/70 nm of Cr/Au, all using poly(meth)acrylate (PMMA) masks patterned by e-beam lithography). Extended Data Fig.~\ref{efig:device_fab}a shows an optical micrograph of the completed device. Contacts are labeled in the image to indicate the specific electrodes used to perform the transport measurements.

\textbf{Transport measurements.} Transport measurements were carried out across two thermal cycles, first in a Bluefors LD dilution refrigerator equipped with a 3-axis superconducting vector magnet and then in a Bluefors XLD dilution refrigerator with a one-axis superconducting magnet. In both systems, the nominal base mixing chamber temperature was $T=10-20$~mK, as measured by a factory-supplied RuO$_x$ sensor. Unless otherwise specified, measurements were carried out at the nominal base temperature of the fridge. Four-terminal lock-in measurements were performed by sourcing a small alternating current of either $I_{ac}=120$~pA or $1.0$~nA at a frequency $<20$ Hz, chosen to accurately capture sensitive transport features while minimizing electronic noise. In addition, a global bottom gate voltage between $-120$~V and $+60$~V was applied to the Si substrate to improve the contact resistance. Data in Figure~\ref{fig:4} was acquired with $I_{ac}=1.0$~nA. Unless otherwise specified, all others were taken with $I_{ac}=120$~pA.

The charge carrier density, $n$, and the out-of-plane electric displacement field, $D$, were defined according to $n= \left(C_{\text{bg}} V_{\text{bg}}+C_{\text{tg}} V_{\text{tg}}\right) / e$ and $D=\left(C_{\text{tg}} V_{\text{tg}} - C_{\text{bg}} V_{\text{bg}}\right) / 2 \epsilon_0$, where $C_{\text{tg}}$ and $C_{\text{bg}}$ are the top and bottom gate capacitance per unit area and $\epsilon_0$ is the vacuum permittivity. $C_{\rm{tg}}$ and $C_{\rm{bg}}$ were estimated by fitting the slopes of the quantum Hall states in the Landau fan shown in Extended Data Fig.~\ref{efig:basic_fans}b.

To reduce geometric mixing between measured longitudinal and transverse voltage when $|B| > 0$, we often plot the field-symmetrized values of resistance. $\rho_{xx}$ was symmetrized according to $\left(\rho_{xx}(B > 0) + \rho_{xx}(B < 0)\right)/2$, and $\rho_{xy}$ was antisymmetrized according to $\left(\rho_{xy}(B > 0) - \rho_{xy}(B < 0)\right)/2$. All of the measurements are (anti-)symmetrized, unless otherwise specified. 

The contacts are labeled in the schematic in Extended Data Fig.~\ref{efig:device_fab}a. Contacts A and B were used as the source and drain electrodes. All measurements of $\rho_{xy}$ were made using contacts 2 and 4. Measurements of $\rho_{xx}$ were made using contacts 1 and 2 (Figs. 1e-i, 2a-d, 3b, and 3e-f), or 2 and 3 (Fig. 1d, 3a, 3c-d, and 4). The RPG/hBN twist angle extracted from the two pairs of voltage probes is identical within experimental resolution, although contact-specific artifacts differ between the pairs (e.g., see Extended Data Fig.~\ref{efig:unsymmetrized_fans}). In all Extended Data and Supplementary Information Figures, we denote which contact pair is being shown. The map in Fig.~\ref{fig:1}b was constructed by summing the signal between contacts 1-2 and 2-3 and dividing by $2$ in order to mitigate contact-specific artifacts. The map was taken piece-wise and stitched together post-measurement, with the voltage on the silicon gate adjusted appropriately for each segment of the map in order to best eliminate contact resistance artifacts. 

\textbf{Determination of moir\'e period and twist angle.} The carrier density, $n_s$, required to fully fill a four-fold degenerate moir\'e band was estimated by fitting the sequence of Brown-Zak oscillations from a Landau fan taken at $D=0$, as shown in Extended Data Fig.~\ref{efig:basic_fans}. The oscillations occur when $\phi/\phi_0=4B/n_s\phi_0=p/q$, where $\phi_0=h/e$ is the magnetic flux quantum and $p$ and $q$ are integers. The value of $n_s$ is confirmed by fitting the sequences of quantum Hall states arising from the fully-filled moir\'e bands (i.e., four electrons or holes per moir\'e unit cell). Using $n_{s}$, the filling factor $\nu$ was defined according to $\nu = n/(n_{s}/4)$. The superlattice density is related to the \moire wavelength by $L_M=\sqrt{8/\sqrt{3}n_s}$, from which we find the twist angle according to the equation \[L_M(\theta, a_\mathrm{G}, \delta_{a}) = \frac{(1 + \delta_{a_\mathrm{G}}) \cdot a}{\sqrt{2 \cdot (1 + \delta_{a}) \cdot \left(1 - \cos\theta\right) + \delta^2}}\] where $a_\mathrm{G} = 0.246~\mathrm{nm}$ is the lattice constant of graphene, $\theta$ is the twist angle between the hBN and the RPG, and $\delta_a=\frac{a_{\mathrm{hBN}}-a_\mathrm{G}}{a_\mathrm{hBN}}\approx0.017$ is the lattice mismatch between the lattice constants of the hBN ($a_\mathrm{hBN}$) and graphene. We note that the conversion to twist angle depends on the choice of the lattice mismatch between graphene and hBN, which is not precisely known. For consistency with prior work~\cite{Lu2024}, we chose a value of 1.7\%. There are also two distinct stacking configurations of RPG and hBN resulting in a long-wavelength moir\'e potential, in which the RPG is aligned with hBN close to either $0^{\circ}$ or $60^{\circ}$~\cite{Kwan2023}. We are unable to distinguish between these possibilities experimentally, or know whether our sample has the same or opposite sense of the alignment compared with that in Ref.~\onlinecite{Lu2024}. All of the calculations are performed assuming a configuration in which the A and B sublattices of the aligned graphene layer reside close to the nitrogen and boron atoms of the hBN, respectively. Despite the inherent ambiguities in extracting the twist angle, we emphasize that there is no ambiguity in extracting the moir\'e period as it can be converted directly from the value of $n_s$ with no free parameters. 

\textbf{Analysis of the full transport map.} Here, we detail the salient features in the transport map shown in Fig.~\ref{fig:1}b. At the charge neutrality point (CNP, $\nu=0$), there is an insulating state surrounding $D=0$ marked by diverging $\rho_{xx}$. This insulator has previously been understood to arise as a consequence of a correlation-driven layer-antiferromagnetic state~\cite{Shi2020, Han2024, Zhang2011, Velasco2012, Weitz2010, Liu2024}. As $|D|$ is raised, the insulator collapses to a semimetallic state and then reemerges as a band insulator due to a single-particle band gap formed at the CNP. At $\nu=\pm4$, there are features with higher resistance than at nearby doping, suggestive of a strongly suppressed density of states or small gaps opening upon fully filling the four-fold (spin and valley) degenerate moir\'e valence and conduction bands. This behavior arises over a small range of $D>0$ for $\nu=-4$, and over a larger range of $D$ for $\nu=+4$. 

Highly resistive states also appear at integer filling factors $\nu=1, 2,$ and $3$ for electron-type doping, as well as at $\nu=-2$ for hole-type doping. These are correlated insulating states, formed as a consequence of strong interactions within the moir\'e flat bands that lift the spin and valley degeneracies. For electron-type doping, these states are more prominent for $D<0$, corresponding to electrons biased towards the moir\'e interface. The corresponding states at $D>0$ are insulating over smaller ranges of $D$, perhaps reflecting the weaker moir\'e potential experienced by electrons biased away from the moir\'e interface. 

There are many other resistive features in the map that drift through the $n-D$ parameter space, with no apparent tie to precise integer $\nu$. These features most likely reflect metallic van Hove singularities (vHs) formed within the moir\'e minibands. These could correspond either to vHs within the non-interacting bands, or to interaction-driven vHs arising when isospin degeneracies are spontaneously broken though a generalized Stoner instability. Many of these features are prevalent in RPG without an hBN moir\'e~\cite{Han2023}. In this sense, the moir\'e potential can be viewed as acting to trigger the formation of insulating states when such vHs features drift close to integer values of $\nu$. Such behavior has been seen previously for aligned vs. misaligned rhombohedral trilayer graphene~\cite{Zhou2021_abctrilayer}.

For $\nu>0$ and large $D<0$, and also for $\nu<0$ and large $D>0$, there are expansive regions of immeasurably large resistance. These regions can sometimes correspond to negative resistance in our four-terminal transport measurements. Such behavior is often seen adjacent to very insulating states in vdW devices, generally corresponding to regimes of percolative transport where the electrical potential between voltage probes is not well defined. There is also an artifact that cuts diagonally through the map, crossing through the insulator at $\nu=-2$ and $D>0$ and intersecting $\nu=D=0$. Unlike other artifacts driven by poor equilibration with the contacts, changing the silicon gate modifies its location in $n-D$ space but does not suppress it completely (see Fig. 1b vs. Fig. S2). We currently do not know the origin of this artifact, but note that it is nearly parallel to the $V_{tg}$ axis and that similar features are routinely seen in $4-6$ layer rhombohedral graphene devices, e.g. Refs.~\onlinecite{Lu2024,Han2024,Sha2024}.

\textbf{First-order phase transition at fractional electron-doped band filling.} An intriguing feature of the Landau fan shown in Fig.~\ref{fig:2}b is the first-order phase transition that curves from $\nu\approx0.2$ near $B=0$ to $\nu\approx0.75$ at $B=5$~T (see Extended Data Fig.~\ref{efig:fan_hysteresis}). This phase transition separates very different quantum Hall states in the $\rho_{xx}$ Landau fan, and at large $B$ corresponds to an abrupt sign reversal in $\rho_{xy}$. There are quantum oscillations on either side of the phase transition with an apparent degeneracy of $1$, indicating that spin and valley are likely polarized in both cases. It is possible that the overdoped side of the phase transition corresponds to the electronic crystal state responsible for the insulator at $\nu=1$, whereas the underdoped side corresponds to a situation in which the crystalline order has collapsed and there is overlap between the lowest and next higher moir\'e conduction band. In this scenario, only a single spin--valley branch of the lowest moir\'e conduction band is filled, whereas the next dispersive band may remain unpolarized. The quantum Hall state arising from $\nu=1$ with Landau level filling factor $\nu_{LL}=-3$ (indicated by the black arrow in Fig. \ref{fig:2}b) provides evidence for this interpretation, as the state appears to be gapped on the overdoped side of the phase transition but gapless on the underdoped side. In the latter case, it is responsible for only weakly modulating the resistance of the partially filled Landau levels emerging from $\nu=0$. More work is needed to better understand the nature of this first-order phase transition and the evolution of the fermiology of surrounding states as a function of doping. However, it is interesting to note that a remarkably similar feature is seen in twisted bilayer-trilayer graphene~\cite{Su2024}, which also hosts TEC states that spontaneously break the discrete translational symmetry of the moir\'e lattice. 

\textbf{Correlated insulators at $\nu=-2$.} In addition to the pocket of valley-polarization near $\nu=-2$ shown in Fig.~\ref{fig:4}, there are also incipient correlated insulating states precisely at $\nu=-2$ over a narrow range of both signs of $D$. These states appear to arise independently from the valley-polarized pocket, despite their close proximity in the $n-D$ map. The nature of these incipient insulators is not entirely clear. The states appear to be trivial insulators, with $C=0$, based on analysis of Landau fan diagrams taken at fixed $D$ (Extended Data Fig.~\ref{efig:mulitband_transport}). The resistance of the states grows rapidly with increasing $B$, but becomes only slightly larger with magnetic field oriented in the 2D plane up to $1$~T. The absence of Chern insulator behavior and associated AHE upon doping appears to be inconsistent with valley polarization. The weakly growing resistance with in-plane field is consistent with spin-polarized insulators, although it appears that the accessible $1$~T range of in-plane field is not sufficient to unambiguously determine the ground state ordering. Curiously, these states arise only over values of $D$ for which the CNP is in a semimetallic state, although it is not clear whether this is meaningful or purely a coincidence. Future work will be needed to better understand the nature of these incipient correlated insulating states, and whether they have any meaningful interplay with the nearby pocket of valley polarization.

\textbf{Modification of Streda formula for gapless states.} As discussed in the main text, the trajectories of the incipient Chern insulator states in the Landau fans in Figs.~\ref{fig:4}f-g are atypical in that they are: (i) occasionally curved, (ii) imply Chern numbers that are larger than the apparent value extracted from the maximum value of $\rho_{xy}$, and (iii) can have anomalously large values as big as $13$. All of these features call into question the applicability of the Streda formula in assigning a Chern number to these states. In particular, there is no simple explanation for the observation of a Streda slope that is larger than the Chern number implied by $\rho_{xy}$, as this behavior cannot easily be ascribed to effects of disorder (which can artificially reduce, but never enhance, the value of the antisymmetrized $\rho_{xy}$ in an incipient Chern insulator). The most notable disagreement lies in the ``primary'' state at $\nu\approx-2.45$, which has an apparent Streda slope of $7$ but $\rho_{xy}\approx h/4e^2$ implying a Chern number of $4$ (see Figs.~\ref{fig:4}b and f). 

However, if there are coexisting itinerant charges residing in additional bands, the system in not in a fully gapped state and the Streda formula is not applicable. In this situation, an additional contribution related to Hall diffusion must be taken into account~\cite{Vignale2020}. In the most general scenario, tracking the evolution of a given topological state necessitates considering how the state evolves with both $B$ and $\mu$, where $\mu$ is the chemical potential: $dn=\left(\frac{\partial n}{\partial B}\right)_\mu dB + \left(\frac{\partial n}{\partial \mu}\right)_B d\mu$. The first term here describes the usual Streda formula, whereas the second arises due to the coexisting itinerant charges. We do not attempt to understand the precise trajectories of the incipient Chern insulators, but speculate that the anomalous slopes of the states we observe may originate from the additional contribution of coexisting itinerant charges in overlapping bands. Such a scenario likely requires that the itinerant charges have low mobility, such that they do not completely suppress the characteristic deep minima in $\rho_{xx}$ and large enhancements of $\rho_{xy}$ arising from the chiral edge modes and incompressible bulk of the Chern insulator formed in the highest moir\'e valence band.

\textbf{Quantum oscillations induced by graphite gates.} Some of the Landau fan diagrams exhibit periodic oscillations as a function of $B$ that do not depend on $\nu$ (e.g., Fig.~\ref{fig:3}b). These are not consistent with Brown-Zak oscillations given the known moir\'e period of the sample. Instead, they appear to arise as a consequence of cyclotron gaps formed on the surface of the graphite gates. Supplementary Information Fig.~S3 shows a Fourier transform analysis of these oscillations for various different Landau fan diagrams. Oscillations are seen across a wide range of $\nu$ and $D$, but all have similar frequencies of $f_B \approx 33$~T. This behavior is most prominent in regions of the Landau fans where the resistance changes rapidly with gate voltage (i.e., large d$R$/d$n$). Collectively, this behavior is consistent with that seen in a previous study of graphite-gated vdW heterostructures, pointing to their origin as oscillations in the graphite density of states~\cite{Zhu2021gates}. Therefore, all such oscillations appearing at fixed $B$ over a wide range of $\nu$ most likely originate from the graphite gates, and do not reflect the intrinsic physics of moir\'e RPG. 

\textbf{Dependence of the transport properties of TEC states on bias current.} The transport properties of the Chern insulators in the electron-doped valley-polarized pocket are extremely sensitive to the bias current. Supplementary Information Fig.~S6 shows measurements of $\rho_{xx}$ and $\rho_{xy}$ taken as a function of dc current bias, $I_{dc}$. To perform these measurements, the dc current is applied in addition to a small ac bias of $14$~pA while measuring the differential resistance, $dV/dI$, with a lock-in amplifier. Supplementary Information Fig.~S6b shows representative line traces taken for the Chern insulators at $\nu=2/3$ and $1$. Both show that $\rho_{xy}$ deviates from its (nearly) quantized value as $I_{dc}$ is raised. Similarly, $\rho_{xx}$ quickly transitions from a deep minimum to a large value close to $h/e^2$. Supplementary Information Fig.~S6c show maps of $\rho_{xx}$ and $\rho_{xy}$ vs. $I_{dc}$ measured across a range of $\nu$. The maps show that both quantities are especially sensitive to even small increases in $I_{dc}$ within the Chern insulator states, but less sensitive in metallic states. 

\textbf{Band structure calculation.} We use the continuum model to calculate the band structure. The Hamiltonian for each valley and spin is:
\begin{equation}
    H_K = H_0 + H_M,
\end{equation}
where $H_0$ is the Hamiltonian of rhombohedral pentalayer graphene:
\begin{equation}
        H_0 = 
    \begin{pmatrix}
    H_1 & \Gamma & \tilde{\Gamma} & \mathbf{0_{2\times 2}} & \mathbf{0_{2\times 2}} \\
    \Gamma^\dagger & H_2 & \Gamma & \tilde{\Gamma} & \mathbf{0_{2\times 2}} \\
     \tilde{\Gamma}^\dagger   & \Gamma^\dagger & H_3 & \Gamma & \tilde{\Gamma}  \\
    \mathbf{0_{2\times 2}} & \tilde{{\Gamma}}^\dagger & \Gamma^\dagger & H_4 & \Gamma \\
   \mathbf{0_{2\times 2}}  & \mathbf{0_{2\times 2}} & \tilde{{\Gamma}}^\dagger & \Gamma^\dagger & H_5
    
    \end{pmatrix},
\end{equation}
with 
\begin{eqnarray}
    &H_i=\begin{pmatrix}
        (\frac{i}{4}-\frac{1}{2})\delta & -\frac{\sqrt{3}}{2}\gamma_0  (\tilde{k}_x-\mathrm{i}\tilde{k}_y)\\
        -\frac{\sqrt{3}}{2}\gamma_0 (\tilde{k}_x+\mathrm{i}\tilde{k}_y) & (\frac{i}{4}-\frac{1}{2})\delta
    \end{pmatrix},&\\
    &\Gamma=\begin{pmatrix}
        -\frac{\sqrt{3}}{2}\gamma_4 (\tilde{k}_x-\mathrm{i}\tilde{k}_y) & -\frac{\sqrt{3}}{2}\gamma_3 (\tilde{k}_x+\mathrm{i}\tilde{k}_y) \\
        \gamma_1 &  -\frac{\sqrt{3}}{2}\gamma_4  (\tilde{k}_x-\mathrm{i}\tilde{k}_y)        
    \end{pmatrix},&\\
    &\tilde{\Gamma}=
    \begin{pmatrix}
        0 & \frac{1}{2}\gamma_2 \\
        0 & 0
    \end{pmatrix}, &
\end{eqnarray}
where the parameters are $(\gamma_0,\gamma_1,\gamma_2,\gamma_3,\gamma_4)=(-2600,358,-8.3,293,144)$~meV. $\delta$ is the potential difference between the top and the bottom graphene layers. $\tilde{k}_x+\mathrm{i}\tilde{k}_y=e^{\mathrm{i}\theta_3}(k_x+\mathrm{i}k_y)$, where $\theta_3=\arctan\frac{\theta}{\delta_a}$. The moir\'e potential is:
\begin{equation}
    H_M(\mathbf{G_j}) = 
    \begin{pmatrix}
        C_0 e^{\mathrm{i}\phi_0}+C_z e^{\mathrm{i}\phi_z} & C_{AB}e^{\mathrm{i}(\frac{(5-j)\pi}{3}-\phi_{AB})}\\
        C_{AB}e^{\mathrm{i}(\frac{(3+j)\pi}{3}-\phi_{AB})} & C_0e^{\mathrm{i}\phi_0}-C_ze^{\mathrm{i}\phi_z}
    \end{pmatrix},
\end{equation}
which represents the tunneling in the first graphene layer aligned with the hBN, with momentum difference given by $\mathbf{G_j}=\frac{4\pi}{\sqrt{3}L_M}(\cos (\frac{j\pi}{3}-\frac{5\pi}{6}),\sin (\frac{j\pi}{3}-\frac{5\pi}{6}))^T$, $j=1,3,5$. For $j=2,4,6$, the tunneling can be determined by taking the Hermitian conjugate. The parameters are determined from DFT calculations~\cite{PhysRevB.108.155406} as $C_0=-10.13~\mathrm{meV}$, $\phi_0=-86.53^\circ$, $C_z=9.01~\mathrm{meV}$, $\phi_z=-8.43^\circ$, $C_{AB}=11.34~\mathrm{meV}$, $\phi_{AB}=-19.60^\circ$. 

\textbf{Hartree--Fock calculations.} We perform Hartree--Fock calculations for $\nu=1$ with a moir\'e period of $11.1$~nm. The BZ used is the moir\'e BZ. The Coulomb interaction term is expressed as: 
\begin{equation}
    H_\mathrm{int}=\frac{1}{2A}\sum_\mathbf{q}\sum_{l,l^\prime}V_{l,l^\prime}(\mathbf{q}):\rho_l(\mathbf{q})\rho_{l^\prime}(\mathbf{-q}):,
\end{equation}
where $\rho_l(\mathbf{q})$ represents the electron density at layer $l$, and $A$ denotes the system area. The interaction potential is written as:
\begin{equation}
    V_{l,l^\prime}(\mathbf{q})=\frac{e^2e^{-q|l-l^\prime|d_\mathrm{layer}}\tanh(q\lambda)}{2\epsilon\epsilon_0 q},
\end{equation}
with $d_\mathrm{layer}=0.34$~nm being the distance between adjacent layers, $\lambda=30$~nm the screening length and $\epsilon=10$ the dielectric constant. We consider four conduction bands in the BZ at the single-particle level and carry out self-consistent calculations. We evaluate 100 random initial ansatzes and choose the one with the lowest energy.

We note that our HF calculations are unable to reproduce the transition between $C=0$ and $C=1$ gapped states at $\nu=1$ for a moir\'e period of $10.8$~nm. Such a transition can be captured either by slightly increasing the moir\'e period to at least $11.3$~nm, or by increasing it to $11.1$~nm and artificially setting the moir\'e potential strength to zero. The origin of this disagreement is currently not clear, and will require further investigation to understand. In order to illustrate a possible physical origin of this topological phase transition, we show calculations in Fig.~\ref{fig:2}e-g in the limit of a vanishing moir\'e potential for a moir\'e period of $11.1$~nm. These calculations are meant to motivate a possible mechanism for generating the transition between $C=0$ and $C=+1$ states at $\nu=1$, rather than perfectly capture the physics of the system. The calculations for the case of the $11.3$~nm moir\'e period and non-zero moir\'e potential are qualitatively consistent with the results shown in Figs.~\ref{fig:2}e-g; thus, the inclusion or exclusion of the weak moir\'e potential does not meaningfully impact the conclusions of this analysis.

\textbf{Spatial charge distribution calculation.} The spatial distribution $n(\mathbf{r})$ is defined as $\langle c^\dagger(\mathbf{r}) c(\mathbf{r})\rangle$, where $c(\mathbf{r})$ is the electron operator in real space. $n(\mathbf{r})$ is calculated by:
\begin{equation}
\begin{split}
    n(\mathbf{r})=&\sum_\mathbf{q}n(\mathbf{q})e^{-\mathrm{i}\mathbf{q\cdot r}}\\
    =&\frac{1}{N_1N_2}\sum_\mathbf{k}\sum_\mathbf{q} \langle u(\mathbf{k+q}) \lvert u(\mathbf{k})\rangle e^{-\mathrm{i}\mathbf{q\cdot r}},
\end{split}
\end{equation}
where $k$ is defined in the BZ, which is discretized into $N_1\times N_2$ points. $\lvert u(\mathbf{k+q})\rangle$ is the periodic part of the Bloch wavefunction. Here, $\mathbf{q}=n_1\mathbf{G_1}+n_2\mathbf{G_2}$ represents any integer linear combination of the crystal reciprocal lattice vectors. The unit of $n(\mathbf{r})$ is the area of the BZ divided by $4\pi^2$.

\textbf{Berry curvature and Chern number calculation.}
We calculate the Berry curvature using the method from Ref.~\onlinecite{fukui2005chern}. The BZ is discretized into an $N_1\times N_2$ grid. For each $\mathbf{k}$ in the BZ, we calculate the $U(1)$ link:
\begin{equation}
    U_\mu (\mathbf{k})=\langle u(\mathbf{k}+\delta k_\mu \mathbf{\hat{k}_\mu})\lvert u(\mathbf{k})\rangle,
\end{equation}
where $\mu=1,2$. The Berry curvature is then given by:
\begin{eqnarray}
    \Omega(\mathbf{k})=\frac{\arg W(\mathbf{k})}{\delta k_1\delta k_2},
\end{eqnarray}
where $W(\mathbf{k})$ is $U_1(\mathbf{k}) U_2(\mathbf{k}+\delta k_1 \mathbf{\hat{k}_1}) U^{-1}_1(\mathbf{k}+\delta k_2 \mathbf{\hat{k}_2}) U_2^{-1}(\mathbf{k})$. We ensure that $-\pi<\arg W(\mathbf{k})\le\pi$. The Chern number is calculated as:
\begin{eqnarray}
    C=\frac{1}{2\pi}\int_{\mathrm{BZ}}d^2\mathbf{k} \Omega(\mathbf{k}).
\end{eqnarray}
In Fig.~\ref{fig:2}g, the flux is calculated as:
\begin{equation}
    \frac{1}{2\pi}\int_{H_{r_k}}d^2\mathbf{k}\Omega(\mathbf{k}),
\end{equation}
where $H_{r_k}$ corresponds to the moir\'e BZ scaled by $r_k$, with $r_k=1$ recovering the original moir\'e BZ.

\textbf{Possible ground state orderings at $\nu=2/3$.} We consider three possible states at $\nu=3$: (I) A FQAH state with the crystal lattice matching the moir\'e wavelength $L_M$; (II) A crystal state with a period $L_\mathrm{crystal}=L_M/\sqrt{2/3}$ in which one folded band is fully filled; (III) A different crystal state with a period $L_\mathrm{crystal}=L_M/\sqrt{1/3}$ in which two folded bands are fully filled. Both (II) and (III) describe states that can be thought of as (generalized) anomalous Hall crystals, differing primarily by the sizes of their unit cells. 

For (I), we first perform a Hartree--Fock calculation at $\nu=1$, followed by an exact diagonalization study by projecting the Coulomb interaction into the lowest HF band~\cite{Zhou2023}. For (II), we perform a Hartree--Fock calculation in a BZ scaled by $\sqrt{2/3}$ compared to the moir\'e BZ, with one electron per unit cell. We manually switch off the moir\'e potential due to its incommensurability with the crystal period. For (III), we perform a Hartree--Fock calculation in a  BZ scaled by $1/\sqrt{3}$ compared to the \moire BZ, with two electrons per unit cell. Supplementary Information Fig.~S13 shows HF band structure calculations illustrating the difference between the TEC states (II) and (III). To the best of our knowledge, these two forms of TEC are indistinguishable in our transport measurements, since both manifest as identical $C=1$ Chern insulators on the device scale. These states could potentially be distinguished by microscopically imaging the electronic crystal owing to their different unit cell sizes. Our HF modeling predicts that states (I), (II), and (III) differ in energy by less than $0.65$~meV per moir\'e unit cell. Therefore, at this point it is not possible to determine the ground state with any realistic degree of confidence. We thus caution that more experimental and theoretical work is needed to fully capture the physics of the phase at $\nu=2/3$.


\section*{Acknowledgements}
The authors thank Long Ju for sharing preliminary data; Hart Goldman, Trithep Devakul, Justin Song, and David Cobden for helpful discussions; Yinong Zhang, Jordan Fonseca, and Jiaqi Cai for technical assistance with the AM-KPFM imaging of graphene stacking domains. Research at the University of Washington on correlation-driven topology in pentalayer graphene was solely supported as part of Programmable Quantum Materials, an Energy Frontier Research Center funded by the U.S. Department of Energy (DOE), Office of Science, Basic Energy Sciences (BES), under award DE-SC0019443. Experiments at the University of British Columbia were undertaken with support from the Natural Sciences and Engineering Research Council of Canada; the Canada Foundation for Innovation; the Canadian Institute for Advanced Research; the Max Planck-UBC-UTokyo Centre for Quantum Materials and the Canada First Research Excellence Fund, Quantum Materials and Future Technologies Program; and the European Research Council (ERC) under the European Union’s Horizon 2020 research and innovation program, Grant Agreement No. 951541. D.W. was supported by an appointment to the Intelligence Community Postdoctoral Research Fellowship Program at University of Washington administered by Oak Ridge Institute for Science and Education through an interagency agreement between the US Department of Energy and the Office of the Director of National Intelligence. M.Y., X.X., and A.O. acknowledge support from the State of Washington-funded Clean Energy Institute. K.W. and T.T. acknowledge support from the JSPS KAKENHI (Grant Numbers 21H05233 and 23H02052) and World Premier International Research Center Initiative (WPI), MEXT, Japan. This work made use of shared fabrication facilities at UW provided by NSF MRSEC 2308979 with graphene device development supported by National Science Foundation (NSF) CAREER award no. DMR-2041972. Theoretical calculations were supported by the National Science Foundation under Grant No. DMR-2237031. This work acknowledges usage of the millikelvin optoelectronic quantum material laboratory supported by the M. J. Murdock Charitable Trust.

\section{Author Contributions} 
A.O. and D.W. developed the sample fabrication capabilities; A.O. fabricated the sample; D.W. and A.O. measured the sample in the Yankowitz lab at UW; R.S. performed follow-up measurements in the Folk lab at UBC, which appear in many of the main text figures, under the supervision of J.F. and in discussion with D.W., A.O., and M.Y.; D.W. and A.O. analyzed the data with the assistance of R.S.; J.Y. developed the AFM-based imaging technique to detect rhombohedral graphene under the supervision of X.X.; M.Y. supervised the project; D.W., A.O, and M.Y. wrote the manuscript with B.Z. and Y.Z. providing theory support; K.W. and T.T. provided the hBN crystals.

\section*{Competing interests}
The authors declare no competing interests.

\section*{Additional Information}
Correspondence and requests for materials should be addressed to Matthew Yankowitz.

\section*{Data Availability}
Source data are available for this paper. All other data that support the findings of this study are available from the corresponding author upon request.

\bibliographystyle{naturemag}
\bibliography{references}

\newpage

\renewcommand{\figurename}{Extended Data Fig.}
\renewcommand{\thesubsection}{S\arabic{subsection}}
\setcounter{secnumdepth}{2}
\setcounter{figure}{0} 
\setcounter{equation}{0}

\onecolumngrid
\newpage
\section*{Extended Data}

\begin{figure*}[ht!]
\includegraphics[width=\textwidth]{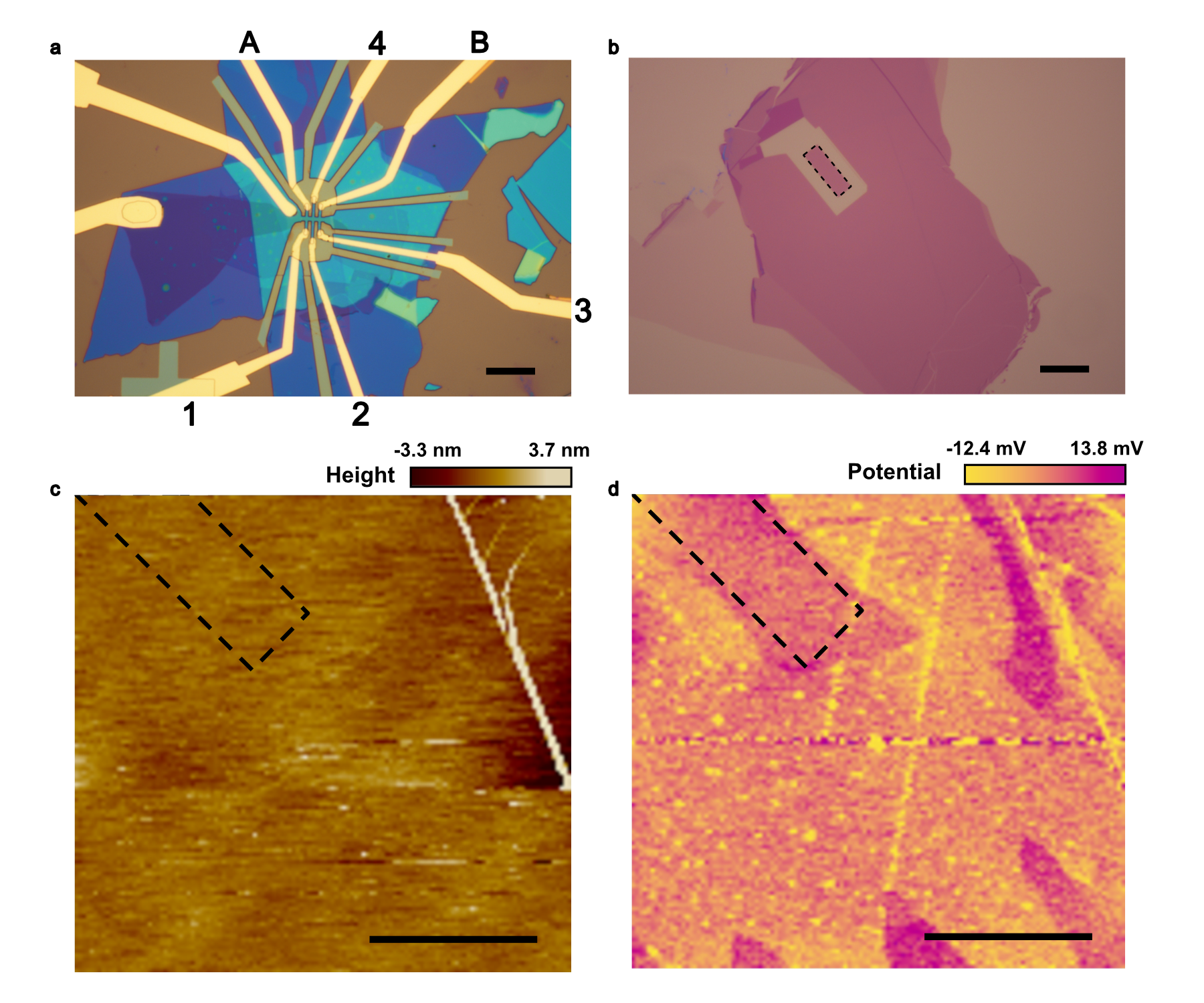} 
\caption{\textbf{Device image and fabrication.} 
\textbf{a}, Optical micrograph of the device. Electrical contacts are labeled A-B and 1-4.
\textbf{b}, Optical micrograph of the pentalayer graphene flake after performing anodic oxidation nanolithography to isolate a rhombohedral domain (outlined by the black dashed rectangle).
\textbf{c}, AFM topograph of the pentalayer flake taken around the area where the anodic oxidation nanolithography was performed (black dashed rectangle). 
\textbf{d}, AM-KPFM measurement of same area as in \textbf{c}. The rhombohedral domain is identified as described in the Methods and in Supplementary Information Fig.~S1. 
All scale bars are 10 $\mu$m. 
}
\label{efig:device_fab}
\end{figure*}

\begin{figure*}[h]
\includegraphics[width=\textwidth]{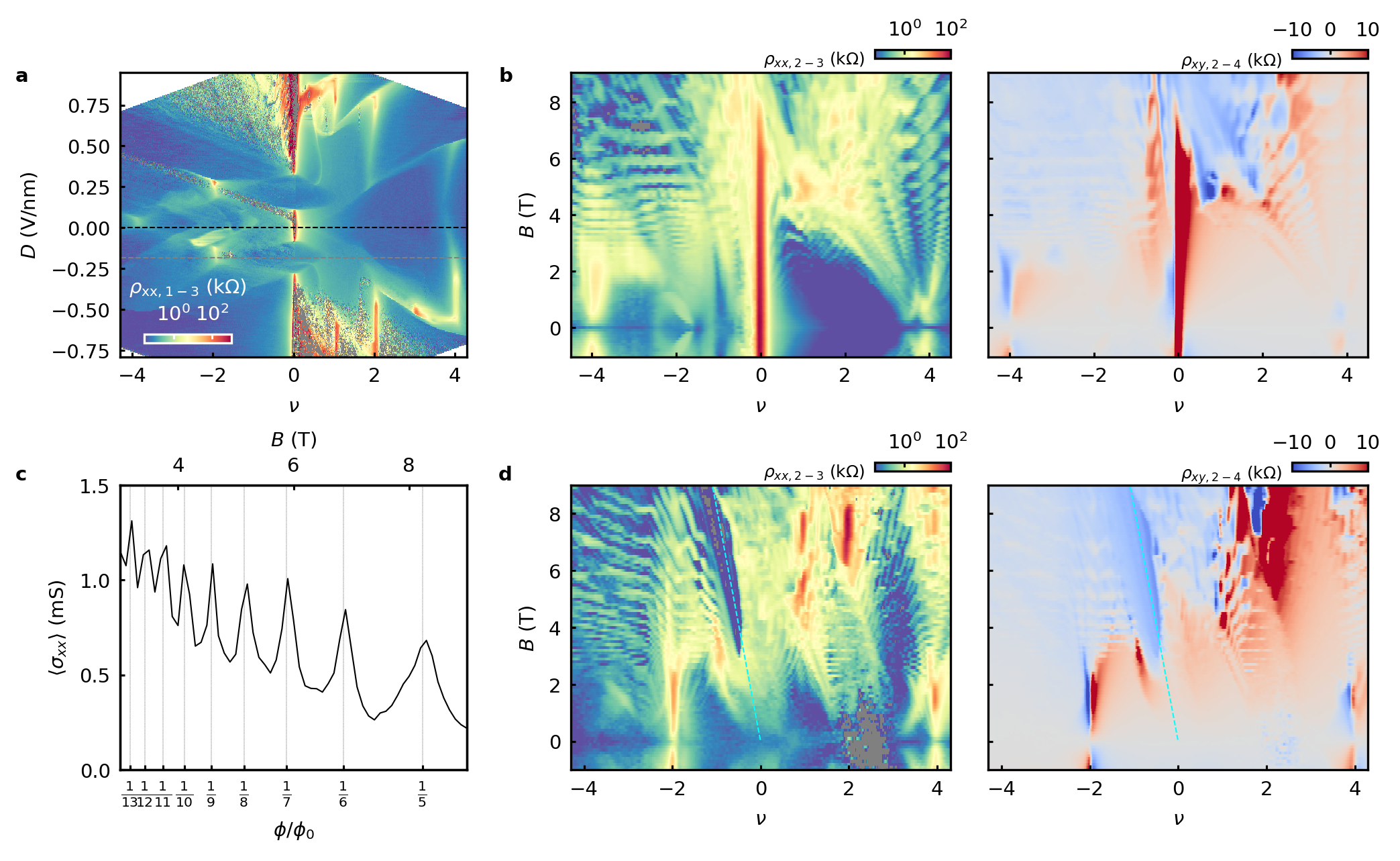} 
\caption{\textbf{Landau fan characterization.}
\textbf{a}, Map of $\rho_{xx}$ at $B=0$ reproduced from Fig.~\ref{fig:1}b, with dashed black and gray lines denoting where the Landau fans in \textbf{b} and \textbf{d} are acquired. 
\textbf{b}, Landau fan of $\rho_{xx}$ and $\rho_{xy}$ acquired at $D=0$ (black line in \textbf{a}). 
\textbf{c}, Averaged conductance, $\langle \sigma_{xx}\rangle = \rho_{xx}/\left(\rho_{xx}^2 + \rho_{xy}^2\right)$ calculated from the Landau fan in \textbf{b} across all accessible $n$ as a function of $B$. Brown-Zak oscillations occur when $\phi/\phi_0=4B/n_s\phi_0=p/q$, where $p$ and $q$ are integers. The best fit to the data yields $n_s=3.98\times10^{12}\ \rm{cm}^{-2}$ (see Methods).
\textbf{d}, Landau fan taken at $D=-0.185\ \rm{V/nm}$ (gray line in \textbf{a}). The blue dashed line denotes a Streda slope of $C=-5$, confirming the sample is rhombohedral pentalayer graphene~\cite{Han2024}. Data in \textbf{b-d} are taken with 1~nA bias current and are not (anti-)symmetrized.
}
\label{efig:basic_fans}
\end{figure*}

\begin{figure*}
\includegraphics[width=6in]{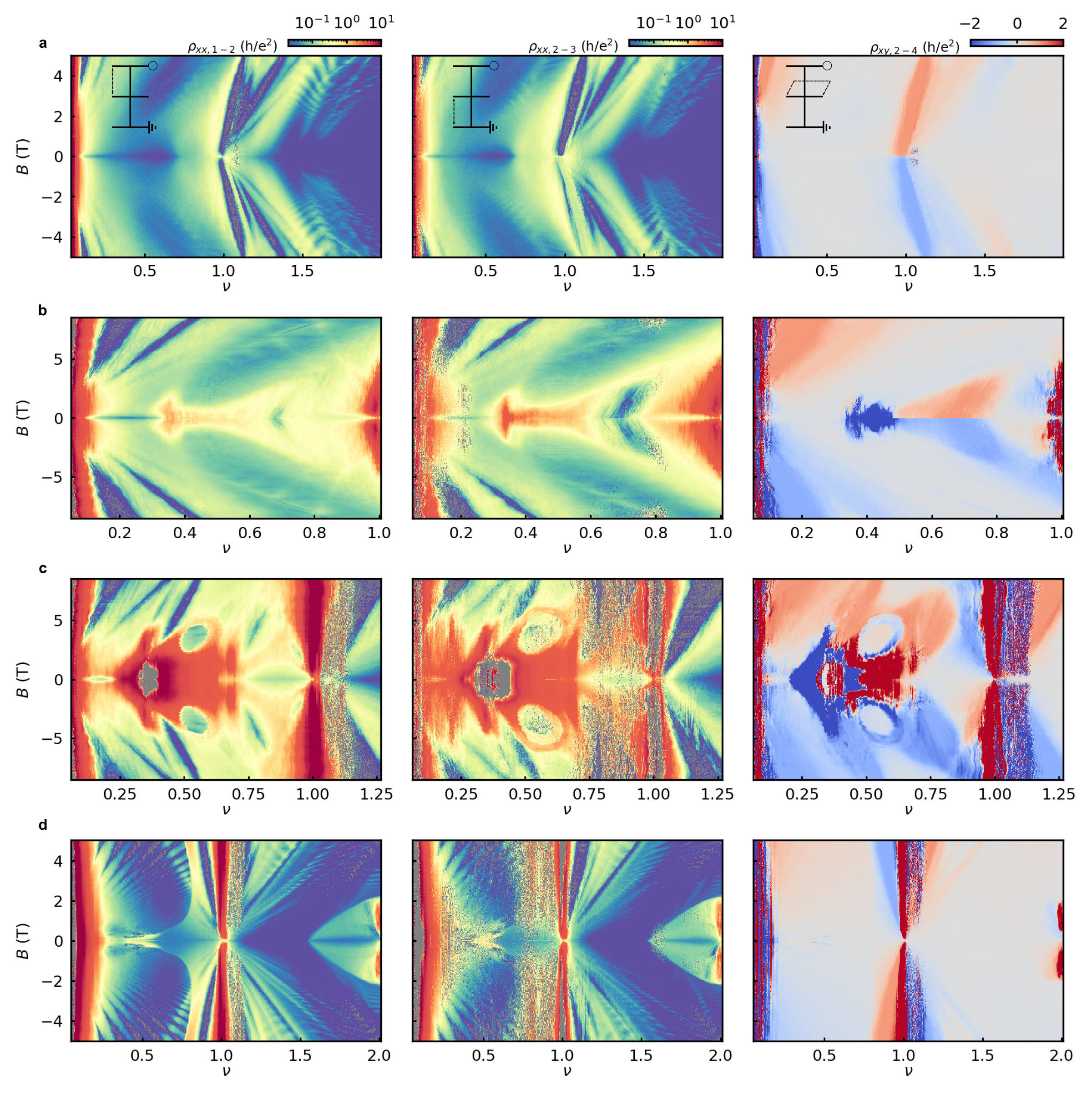} 
\caption{\textbf{Contact pair variation and unsymmetrized Landau fans.}
Raw data (i.e., without symmetrization and antisymmetrization) for the Landau fans taken with different contact pairs at \textbf{a}, $D=0.91\ \rm{V/nm}$, \textbf{b}, $D=0.8625\ \rm{V/nm}$, \textbf{c}, $D=0.826\ \rm{V/nm}$, and \textbf{d}, $D=0.740\ \rm{V/nm}$. Schematics in \textbf{a} denote which contact pairs were used for the measurements, where current was sourced at the top right (circle) and drained to the bottom right (ground) and the potential difference was measured between the contact pairs denoted by the dashed black arrows. The device had two $\rho_{xx}$ contact pairs (leftmost and center plots, using the same color scale) and one $\rho_{xy}$ contact pair (rightmost plots). The two $\rho_{xx}$ contact pairs are very similar overall, although exhibit small differences in the detailed features of states indicating slight sample inhomogeneity. All three contact pairs exhibit vertical strips where the data appears ``speckled''; these are artifacts resulting from poor contact equilibration for certain ranges of $\nu$. We were unable to eliminate these artifacts using the silicon gate voltage.
}
\label{efig:unsymmetrized_fans}
\end{figure*}

\begin{figure*}
\includegraphics[width=\textwidth]{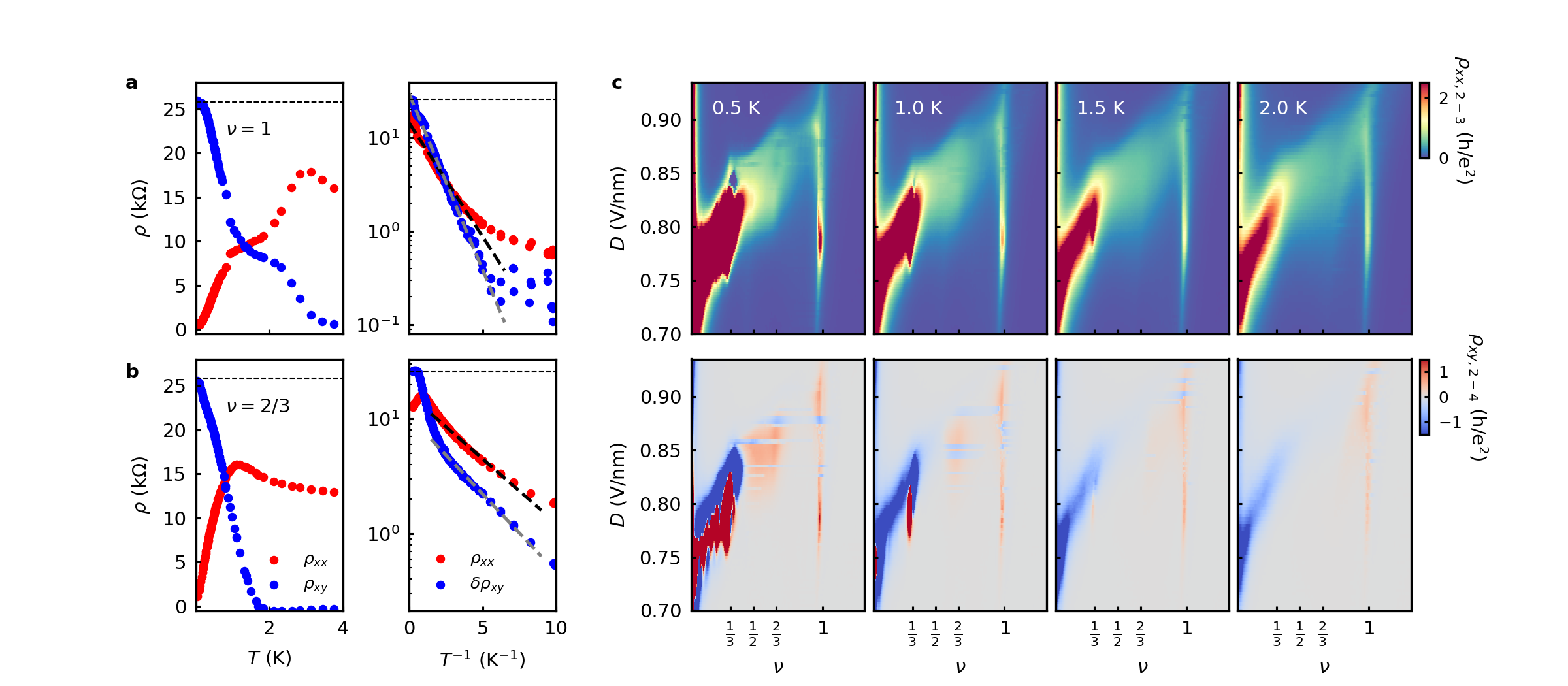} 
\caption{\textbf{Temperature dependence of the correlated states at $0 < \nu \lessapprox 1$.}
\textbf{a}, $\rho_{xx}$ and $\rho_{xy}$ measured as a function of temperature at $\nu=1$ and $D=0.909$~V/nm. Dashed black lines indicate $\rho=h/e^2$. The right panel is shown as an Arrhenius plot, from which we extract gap sizes based on the slopes in the thermally activated regime. Here, $\delta \rho_{xy}=\rho_{xy}[T=20\ \mathrm{mK}]-\rho_{xy}$. The dashed black line is the fit to the $\rho_{xx}$ curve, assuming $\rho_{xx}\propto \exp{\left(-\Delta/2k_BT\right)}$, and yields $\Delta_{\nu=1}^{xx}=0.096\pm 0.001\ \rm{meV}$. The dashed gray line is a fit to the $\delta \rho_{xy}$ curve, which yields $\Delta_{\nu=1}^{xy}=0.149\pm 0.001\ \rm{meV}$.
\textbf{b}, Similar plots for the $\nu=2/3$ state at $D=0.870$~V/nm. Here, the gap size is determined to be $\Delta_{\nu=2/3}^{xx}=0.044\pm 0.001\ \rm{meV}$ and $\Delta_{\nu=2/3}^{xy}=0.054\pm 0.001\ \rm{meV}$.
\textbf{c}, $\rho_{xx}$ and $\rho_{xy}$ maps at $B=0$ in $0.5$~K increments, as indicated in the top panel. 
Data taken with $I_{ac}=14~\rm pA$ in \textbf{a-b} and $I_{ac}=1~\rm nA$ in \textbf{c}.
}
\label{efig:AHC_temp_dependence}
\end{figure*}

\begin{figure*}
\includegraphics[width=6in]{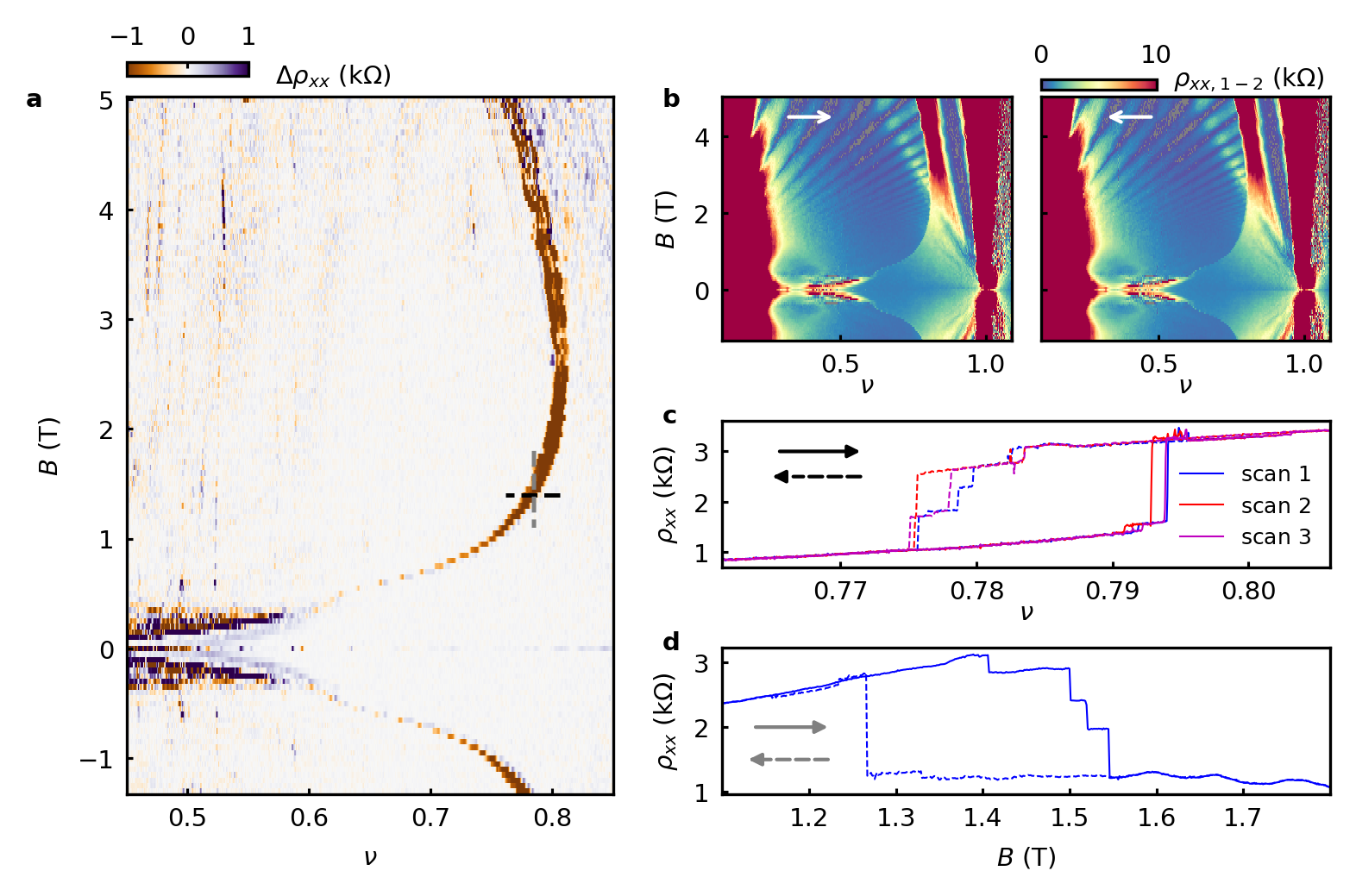} 
\caption{\textbf{Hysteresis across the sharp curved feature in the $D=0.740$~V/nm Landau fan.}
\textbf{a}, Map of the difference between the resistance measured upon sweeping the doping back and forth as a function of magnetic field, $\Delta \rho_{xx}$. There is substantial hysteresis along a curved trajectory matching the abrupt phase transition in the Landau fan in Fig.~\ref{fig:2}b, indicating that it is a first-order phase transition. As argued in the text, the right-hand side of the phase transition appears to be a state with electronic crystalline order, whereas the left-hand side appears to correspond to a state without spontaneous translational symmetry breaking.
\textbf{b}, Raw data of the Landau fans acquired with each sweeping direction used to construct the map in \textbf{a}.
\textbf{c}, Three consecutive measurements (scans 1-3) of $\rho_{xx}$ taken as the doping is swept back and forth at fixed $B$ denoted by the horizontal black dashed line in \textbf{a}. The hysteresis with doping and non-reproducibility upon repeated scanning are all indications of the first-order phase transition. 
\textbf{d}, A similar measurement at fixed $\nu$ but instead sweeping $B$ back and forth, taken along the path indicated by the vertical gray dashed line in \textbf{a}. The measurement also exhibits substantial hysteresis.
}
\label{efig:fan_hysteresis}
\end{figure*}

\begin{figure*}
\includegraphics[width=\textwidth]{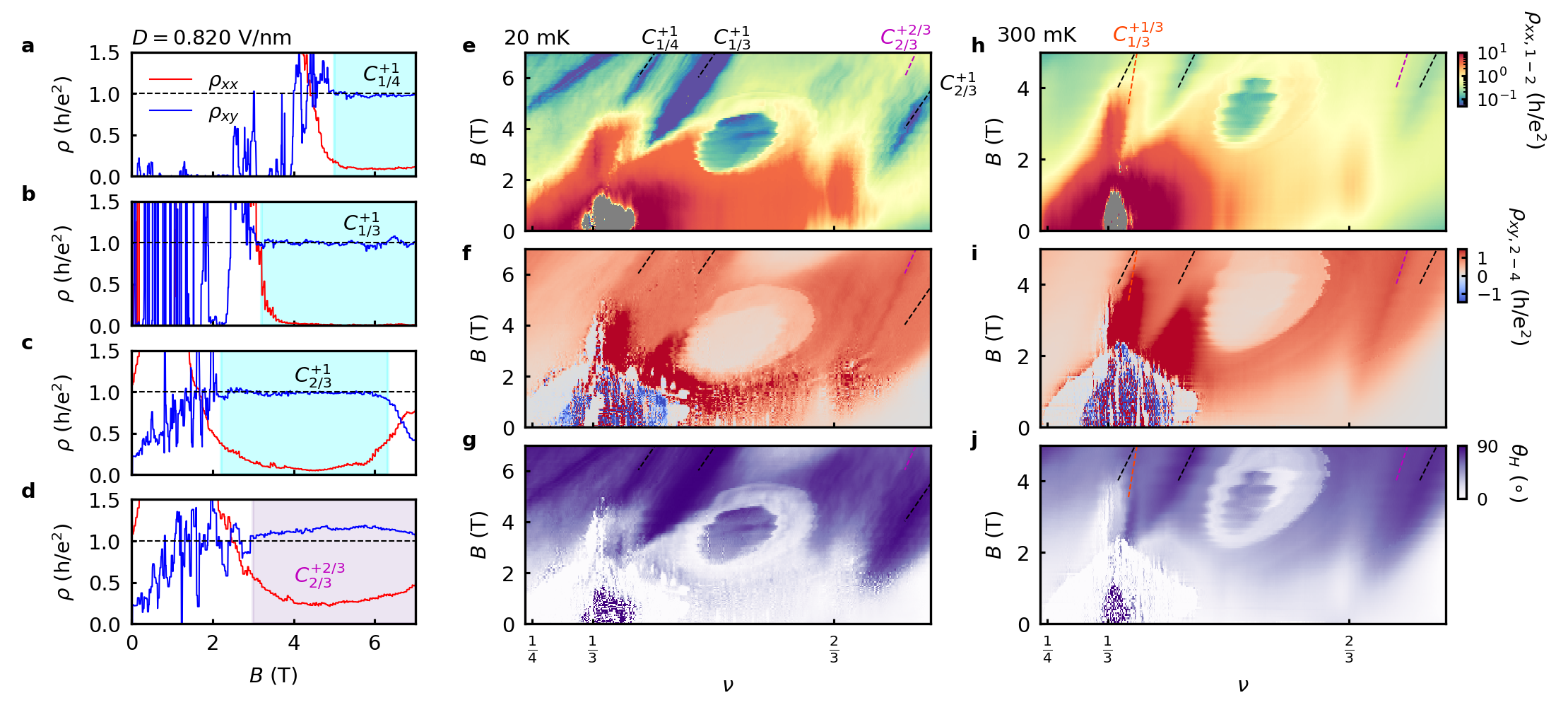} 
\caption{\textbf{Line cuts and temperature dependence of the $D=0.820$~V/nm Landau fan.}
\textbf{a}, Line traces of $\rho_{xx}$ and $\rho_{xy}$ taken along the trajectory indicated by the black dashed line associated with the $C^{+1}_{1/4}$ state in \textbf{e}.
\textbf{b}, Same for the $C^{+1}_{1/3}$ state.
\textbf{c}, Same for the $C^{+1}_{2/3}$ state.
\textbf{d}, Same for the $C^{+2/3}_{2/3}$ state corresponding to the trajectory indicated by the purple dashed line in \textbf{e}.
\textbf{e}, Landau fan of $\rho_{xx}$ acquired at $20$~mK.
\textbf{f}, Same for $\rho_{xy}$. Note that \textbf{c}-\textbf{f} are reproduced from Fig.~\ref{fig:3} for completeness.
\textbf{g}, Landau fan showing the Hall angle, $\theta_H$. Regions of darkest purple are consistent with Chern insulators.
\textbf{h-j}, Similar measurements as in \textbf{e-g} but at $300$~mK and shown over a slightly different range of $B$ and $\nu$. The orange dashed line marks the trajectory of a $C^{+1/3}_{1/3}$ FCI state. There is a simultaneous dip in $\rho_{xx}$, a large enhancement of $\rho_{xy}$, and a large enhancement of $\theta_H$ over an intermediate range of magnetic field ($2$~T $\lessapprox B \lessapprox 4$~T) in these measurements, potentially indicating an incipient in-field FCI state corresponding to $\nu=1/3$. 
}
\label{efig:fan_with_Hall_angles}
\end{figure*}

\begin{figure*}
\includegraphics[width=\textwidth]{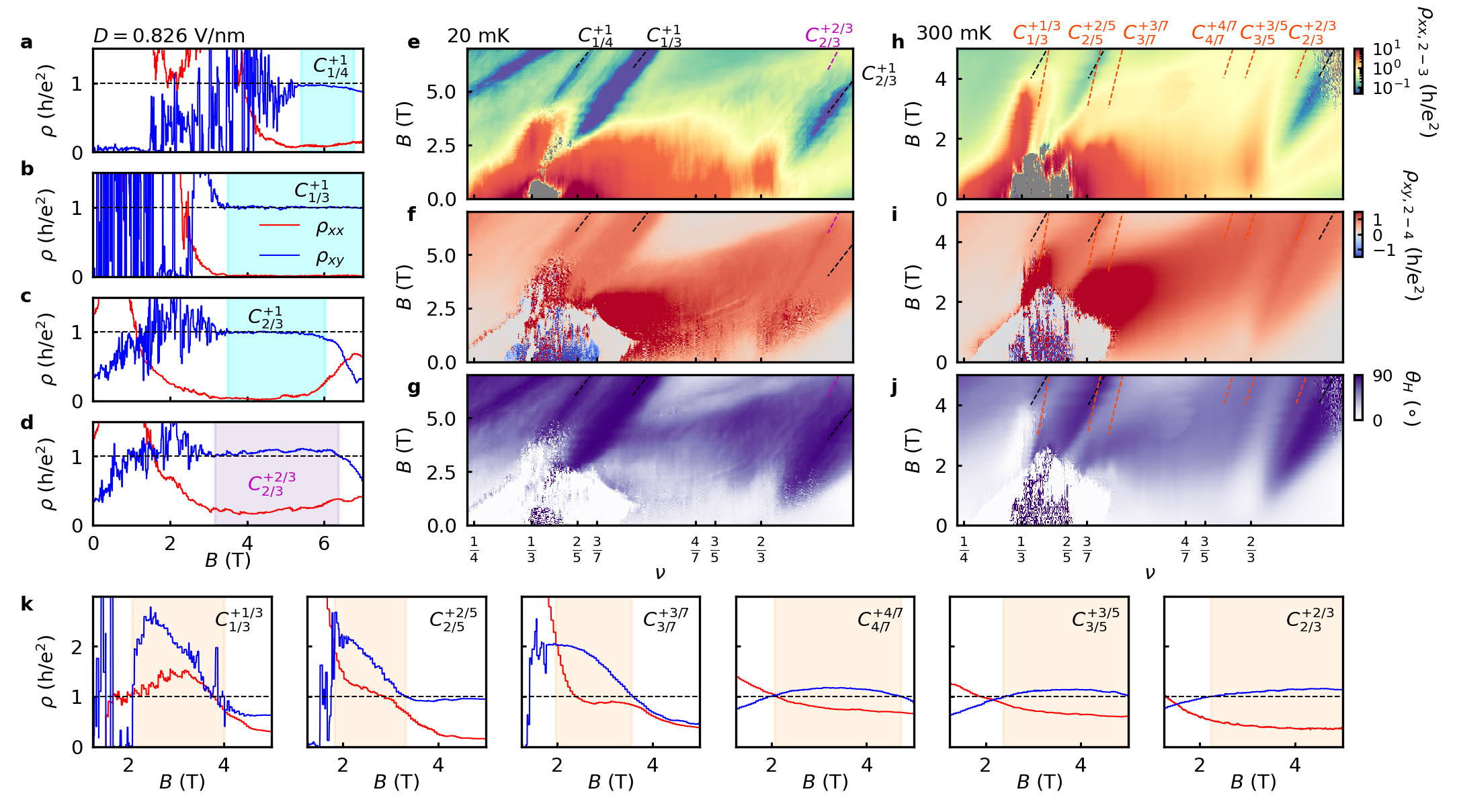} 
\caption{\textbf{Line cuts and temperature dependence of a $D=0.826$~V/nm Landau fan.}
\textbf{a-j}, Same plots for the Landau fan taken at a value of $D$ that is only $6$~mV/nm larger than in Extended Data Fig.~\ref{efig:fan_with_Hall_angles}. Despite this small difference, a number of features change. Perhaps most notably, the oval near $\nu \approx 0.6$ in which there is a first-order phase transition to a different state is gone (except for a weak hint of its recurrence in the fan at $300$~mK). Additionally, the $300$~mK fan has additional faint features possibly indicating other in-field FCI states. There are enhanced regions of $\theta_H$ for $2$~T $\lessapprox B \lessapprox 3$~T consistent with the anticipated slopes of FCI states corresponding to $\nu=2/5$ and $3/7$. There is also a weak feature for $2$~T $\lessapprox B \lessapprox 5$~T along the trajectory of an FCI state corresponding to $\nu=3/5$.
\textbf{k} Line traces of $\rho_{xx}$ and $\rho_{xy}$ at $T=300$~mK taken along the trajectories of all the Jain-sequence states up to denominator of $7$. The orange shaded regions correspond to contiguous regions satisfying both $\rho_{xy}>h/e^2$ and $\rho_{xy} > \rho_{xx}$. We stress that these observations alone do not prove the existence of these additional Jain-sequence in-field FCI states, but motivate their possible emergence. 
}
\label{efig:other_fan_with_Hall_angles}
\end{figure*}

\begin{figure*}
\includegraphics[width=5in]{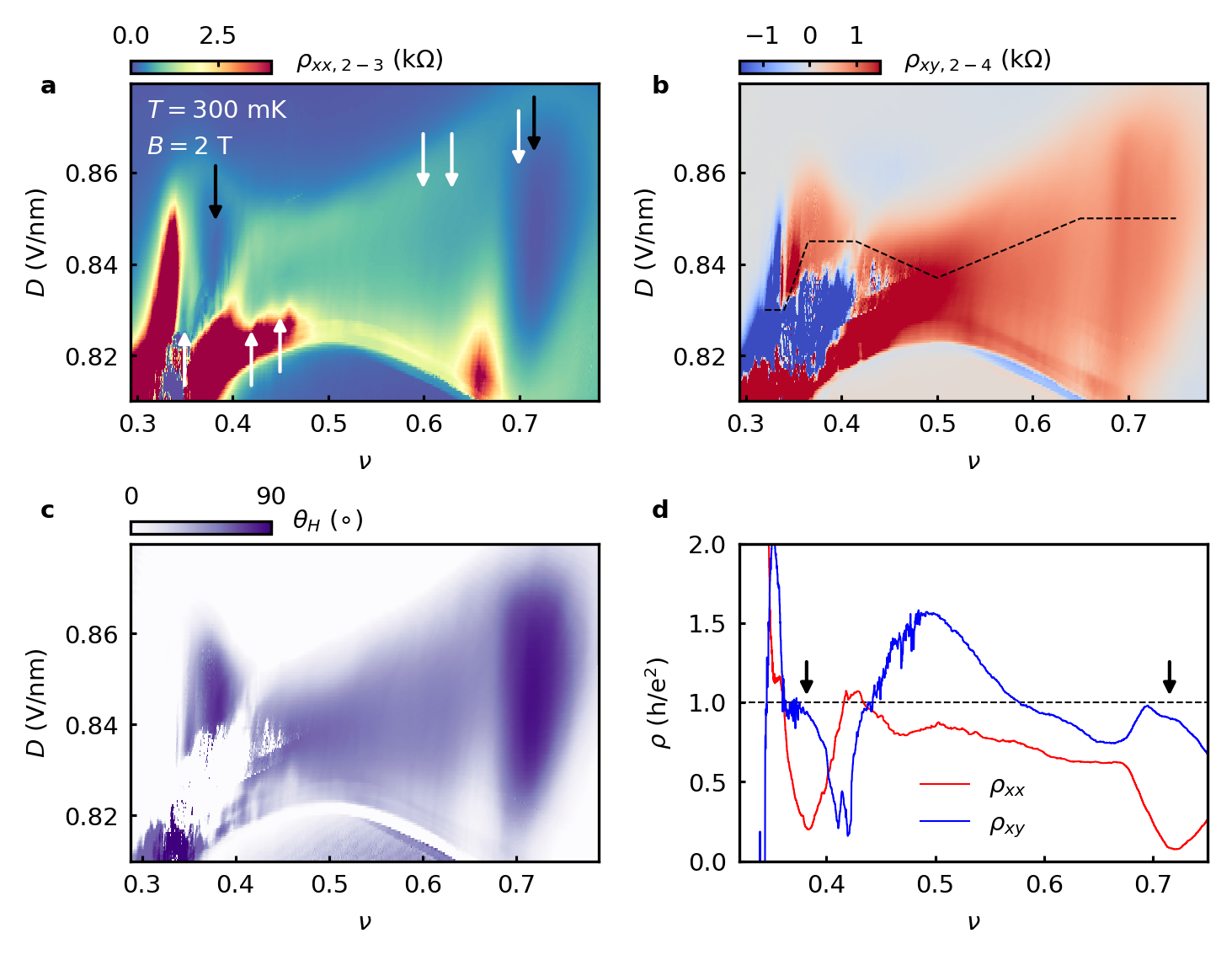} 
\caption{\textbf{Intermixed FCI and TEC states at $B=2$~T.}
\textbf{a}, Map of $\rho_{xx}$ symmetrized at $|B|=2$~T and $T=300~\rm mK$. The black arrows indicate the anticipated positions of $C=+1$ states originating from $\nu=1/3$ and $2/3$, based on their evolution in $B$ as described by the Streda formula. The white arrows correspond to the anticipated positions of Jain-sequence FCI states, from left to right: $\nu=1/3$, $2/5$, $3/7$, $4/7$, $3/5$, $2/3$. These markers are shown only to indicate the anticipated positions of these states, and do not necessarily imply that there are features there (although in many cases there do appear to be). 
\textbf{b}, Antisymmetrized map of $\rho_{xy}$. The curved feature at the bottom of the maps marks the first-order phase transition analyzed in Extended Data Fig.~\ref{efig:fan_hysteresis}.
\textbf{c}, Map of $\theta_H$.
\textbf{d}, Line traces of $\rho_{xx}$ and $\rho_{xy}$ taken along the black dashed path shown in \textbf{b}. The black arrows denote the positions of the $C^{+1}_{1/3}$ (left) and $C^{+1}_{2/3}$ (right) states based upon their evolution described by the Streda formula. Nearby these positions, $\rho_{xy} \approx h/e^2$ and there is a deep suppression of $\rho_{xx}$, consistent with the $C=+1$ Chern insulators analyzed in the main text. In between, there is a wide range of fractional $\nu$ where the antisymmetrized $\rho_{xy}$ far exceeds $h/e^2$. Although there is also a very large $\rho_{xx}$ approaching $h/e^2$, there is nevertheless a large Hall angle. Together with the analysis of the Landau fans in Extended Data Figs.~\ref{efig:fan_with_Hall_angles} and~\ref{efig:other_fan_with_Hall_angles}, this behavior is possibly consistent with in-field FCI states. The large $\rho_{xx}$ has been seen in other FCI states reported in Refs.~\onlinecite{Lu2024} and~\onlinecite{Xie2024FQAH}.
}
\label{efig:2T_map}
\end{figure*}

\begin{figure*}
\includegraphics[width=7in]{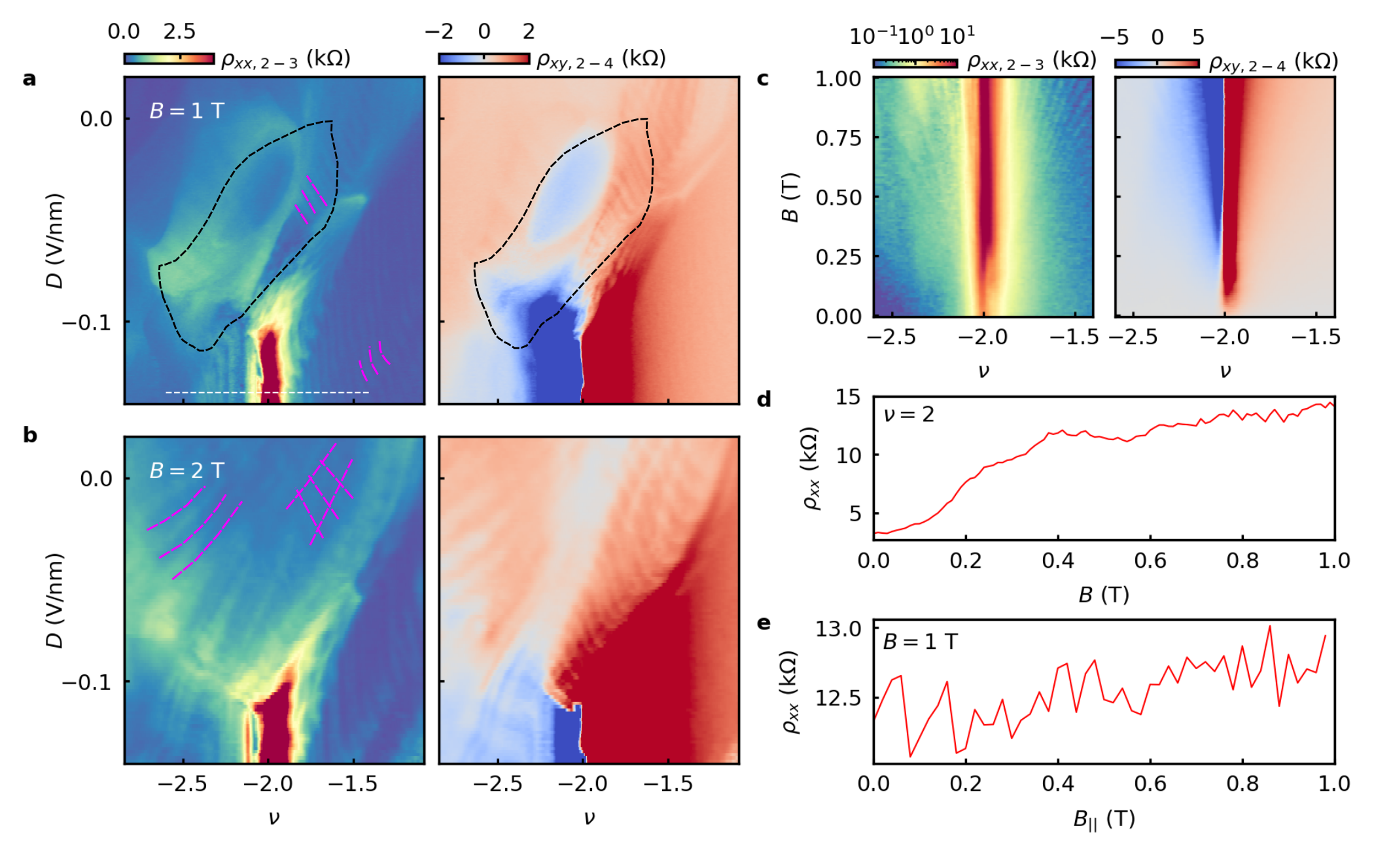} 
\caption{\textbf{Multiband transport and correlated insulator near $\nu=-2$.}
\textbf{a}, Maps of $\rho_{xx}$ (left) and $\rho_{xy}$ (right) taken at $B=1$~T. The black dashed enclosure denotes the boundaries of the valley-polarized pocket at $B=0$ (corresponding to the condition of $\rho_{xy}=0$). In general, quantum oscillations do not remain at fixed values of $\nu$ as $D$ is changed (selected examples are denoted by the dashed purple curves). Such behavior indicates that these states originate from a Fermi surface with multiple coexisting pockets. 
\textbf{b}, The same maps from \textbf{a}, taken at $B=2$~T. The valley-polarized pocket is completely gone at this magnetic field. There are crisscrossing quantum oscillations within the region of $n-D$ occupied by the valley-polarized pocket at $B=0$, suggesting that this pocket arises out of a Fermi surface with multiple pockets.
\textbf{c}, Landau fan of $\rho_{xx}$ (left) and $\rho_{xy}$ (right) taken at $D=-0.135$~V/nm (dashed white line in \textbf{a}). The primary state is a trivial $C=0$ insulator at $\nu=-2$, consistent with a spin-polarized insulator. 
\textbf{d}, $\rho_{xx}$ measured as a function of $B$ at $\nu=2$.
\textbf{e}, The same measurement performed at $B=1$~T as a function of in-plane magnetic field, $B_{||}$. All data taken with 1~nA bias current.
}
\label{efig:mulitband_transport}
\end{figure*}

\begin{figure*}
\includegraphics[width=\textwidth]{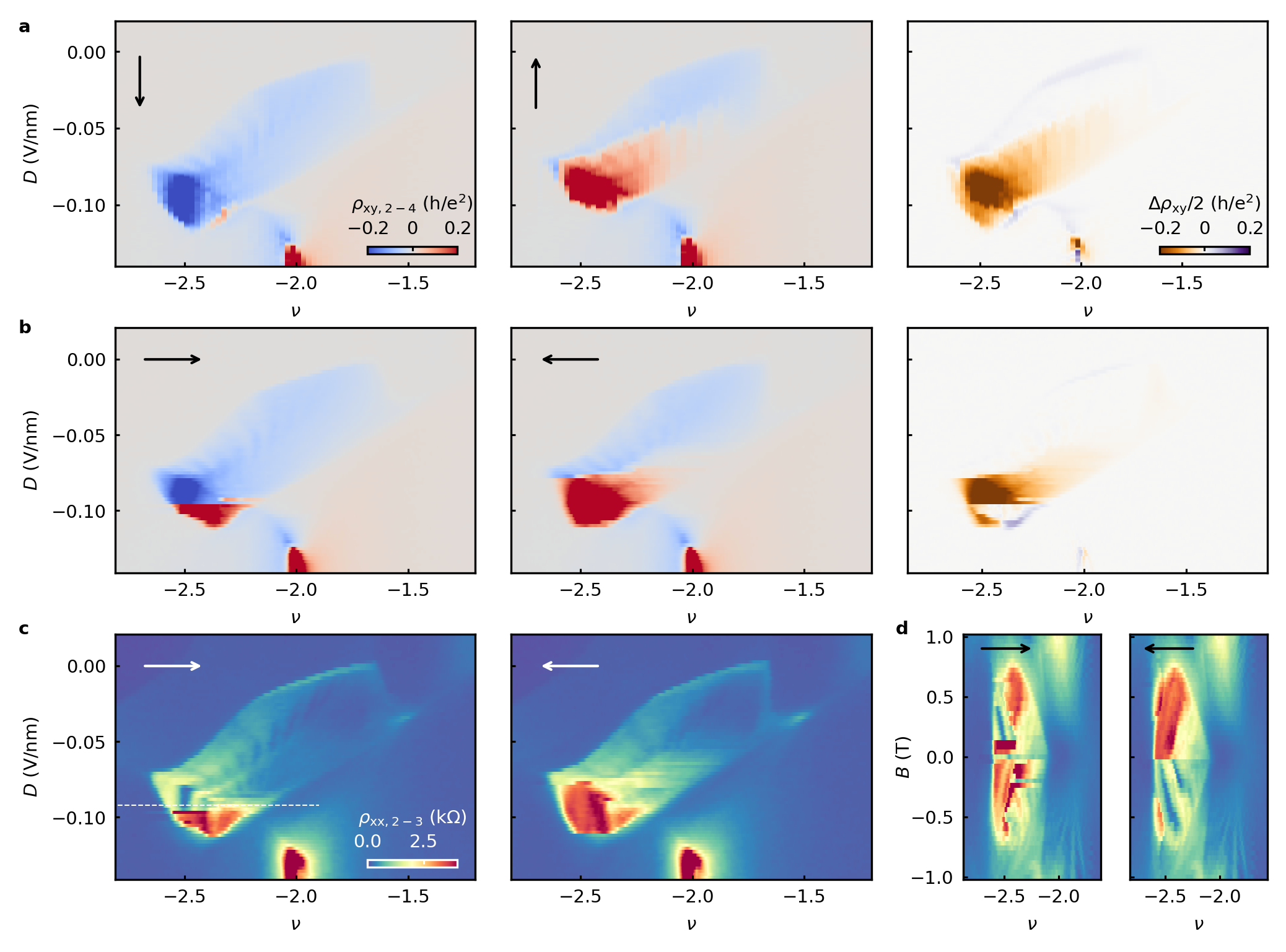} 
\caption{\textbf{Hysteresis of the orbital magnetic state and incipient Chern insulators near $\nu=-2$.} 
\textbf{a}, Map of $\rho_{xy}$ showing the hole-doped valley-polarized pocket, acquired by sweeping $D$ down (left) and up (middle) as the fast axis with $\nu$ as the slow axis. (right) Difference between the down and up sweeps, $\Delta \rho_{xy}/2$.
\textbf{b}, Same as \textbf{a}, but with $\nu$ as the fast axis and $D$ the slow axis.
\textbf{c}, $\rho_{xx}$ maps corresponding to those shown in \textbf{b}.
Data in \textbf{a-c} were taken at $B=0.2~\rm T$ (not symmetrized).
\textbf{d}, Landau fans taken along the white dashed line in \textbf{c}, sweeping the density axis forward and backward. The sign of the slopes of the most prominent Chern states switch upon changing the sweeping direction, reflecting the flipping sign of the orbital magnetism. All data taken with 1~nA bias current.
}
\label{efig:orbital_magnet}
\end{figure*}

\end{document}



\title{Supplementary Information: Interplay of electronic crystals with integer and fractional Chern insulators in moir\'e pentalayer graphene}

\author{Dacen Waters$^{1,2*}$}
\author{Anna Okounkova$^{1*}$}
\author{Ruiheng Su$^{3,4}$}
\author{Boran Zhou$^{5}$}
\author{Jiang Yao$^{1}$}
\author{Kenji Watanabe$^{6}$}
\author{Takashi Taniguchi$^{7}$}
\author{Xiaodong Xu$^{1,8}$}
\author{Ya-Hui Zhang$^{5}$}
\author{Joshua Folk$^{3,4}$}
\author{Matthew Yankowitz$^{1,8\dagger}$}

\affiliation{$^{1}$Department of Physics, University of Washington, Seattle, Washington, 98195, USA}
\affiliation{$^{2}$Intelligence Community Postdoctoral Research Fellowship Program, University of Washington, Seattle, Washington, 98195, USA}
\affiliation{$^{3}$Quantum Matter Institute, University of British Columbia, Vancouver, British Columbia, V6T 1Z1, Canada}
\affiliation{$^{4}$Department of Physics and Astronomy, University of British Columbia, Vancouver, British Columbia, V6T 1Z1, Canada}
\affiliation{$^{5}$Department of Physics and Astronomy, Johns Hopkins University, Baltimore, Maryland, 21205, USA}
\affiliation{$^{6}$Research Center for Electronic and Optical Materials, National Institute for Materials Science, 1-1 Namiki, Tsukuba 305-0044, Japan}
\affiliation{$^{7}$Research Center for Materials Nanoarchitectonics, National Institute for Materials Science, 1-1 Namiki, Tsukuba 305-0044, Japan}
\affiliation{$^{8}$Department of Materials Science and Engineering, University of Washington, Seattle, Washington, 98195, USA}
\affiliation{$^{*}$These authors contributed equally to this work.}
\affiliation{$^{\dagger}$myank@uw.edu (M.Y.)}

\maketitle

\begin{figure*}
\includegraphics[width=\textwidth]{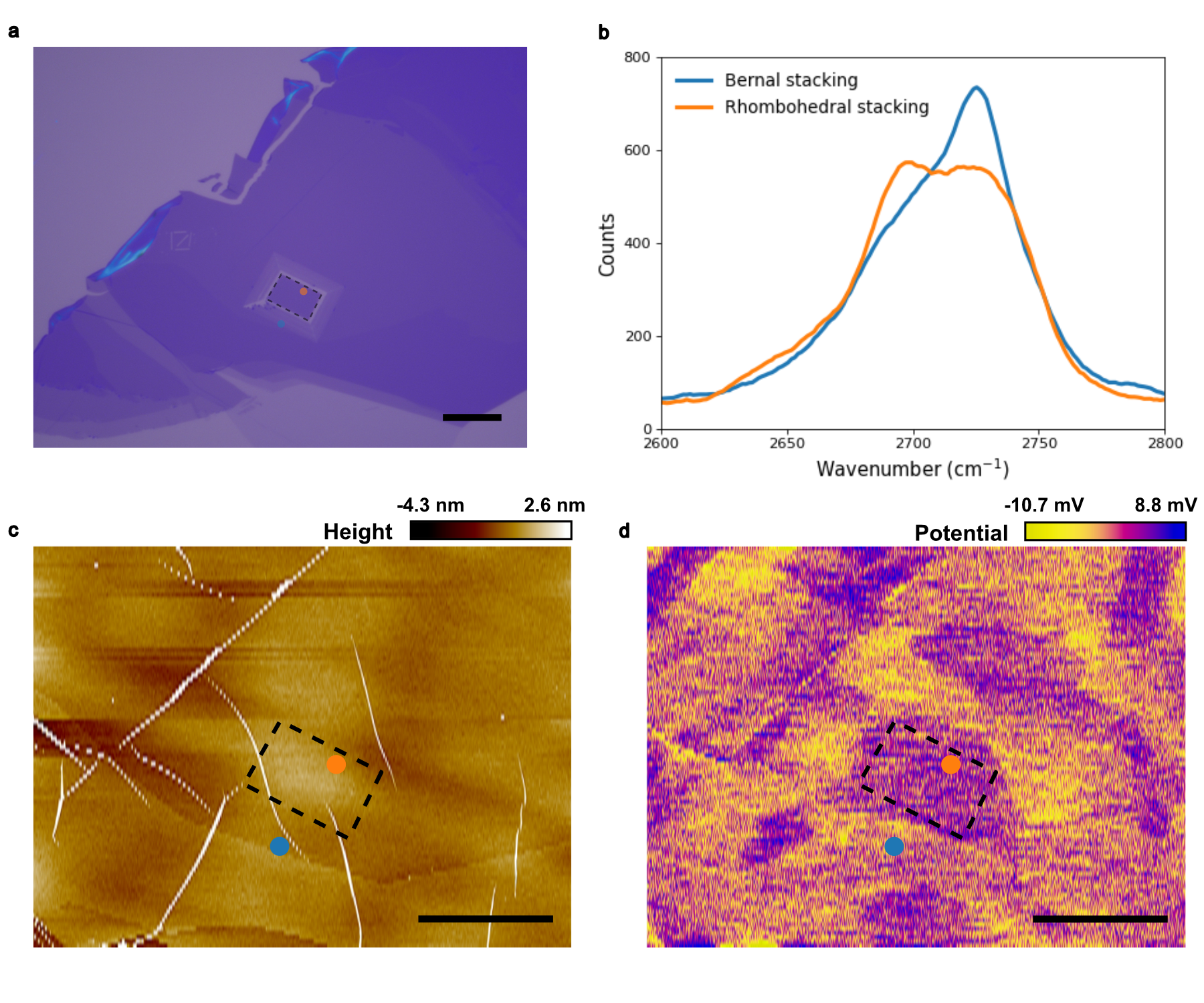} 
\caption{\textbf{Characterization and isolation of rhombohedral domains using AFM and Raman spectroscopy.} 
\textbf{a}, Optical micrograph of a pentalayer graphene flake with a region of anodic oxidation nanolithography etching outlined by the black dashed box. The blue and orange dots correspond to locations where Raman spectroscopy was performed. 
\textbf{b}, Raman spectra corresponding to the positions denoted in \textbf{a}. 
\textbf{c}, AFM topograph of the pentalayer graphene flake (before etching), with the same black box and blue/orange markers as in \textbf{a}. 
\textbf{d}, AM-KPFM scan of same area as in \textbf{c}. 
Scale bars in \textbf{a}, \textbf{c}, and \textbf{d} are all 10 $\mu$m. 
}
\label{sfig:orbital_magnet}
\end{figure*}

\begin{figure*}
\includegraphics[width=5in]{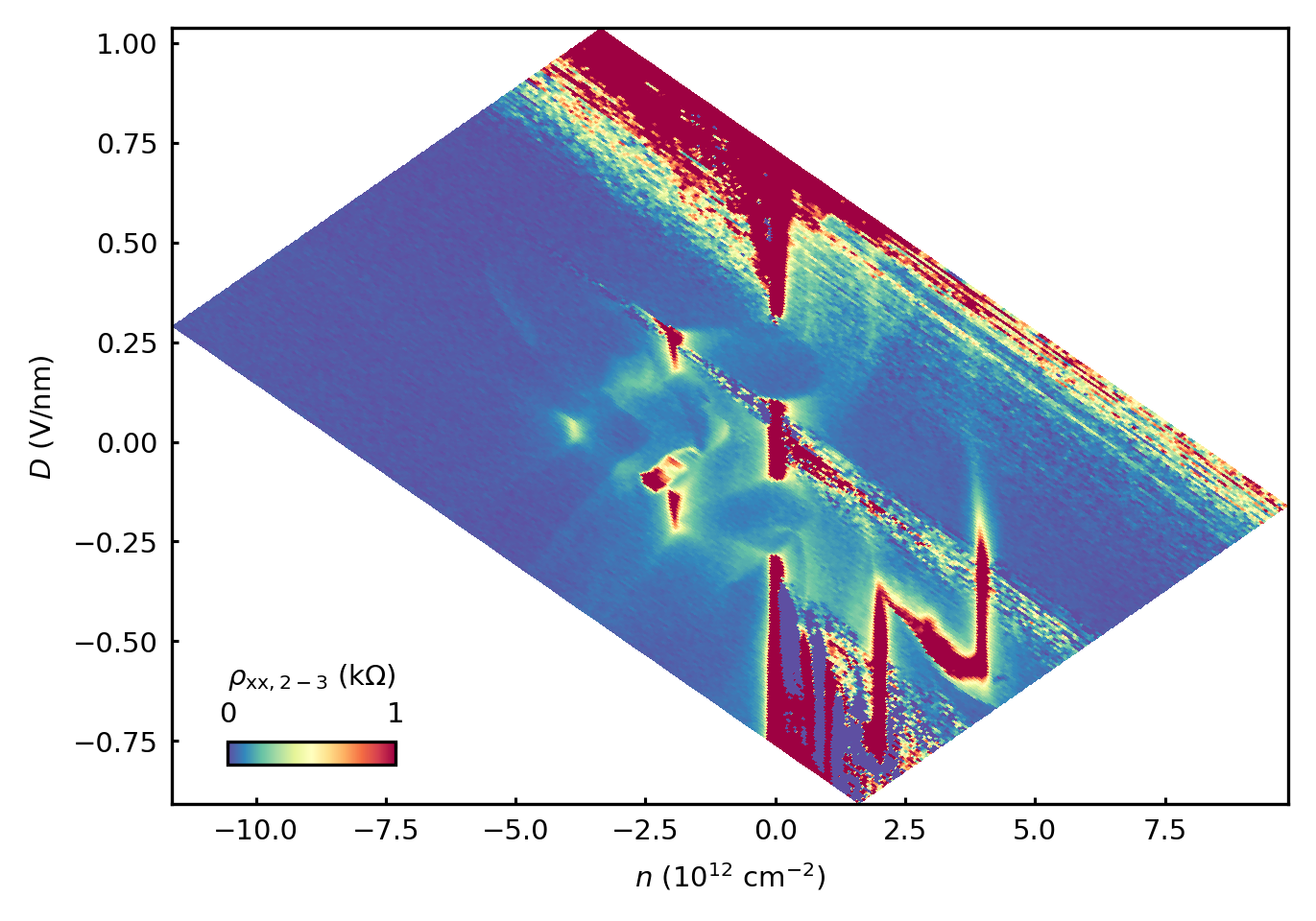} 
\caption{\textbf{Large density range transport map.}
Map of $\rho_{xx}$ taken to the largest experimentally accessible value of $n$. Besides those analyzed in the main text, we observe no additional superlattice features out to $\left| n \right|=11.59\times 10^{12}\ \rm cm^{-2}$. This sets an upper limit on a secondary \moire wavelength from the misaligned hBN of $6.31\ \rm nm$. Assuming a lattice mismatch of $1.7\%$, this corresponds to a lower limit on the twist angle of the misaligned hBN of $2.03^\circ$. The streaking on the upper right hand side of the map is a contact issue, which we can tune away using an appropriate value of the silicon gate voltage.
Data taken with $I_{ac}=1~\rm nA$.}
\label{sfig:large_density_range_map}
\end{figure*}

\begin{figure*}
\includegraphics[width=6in]{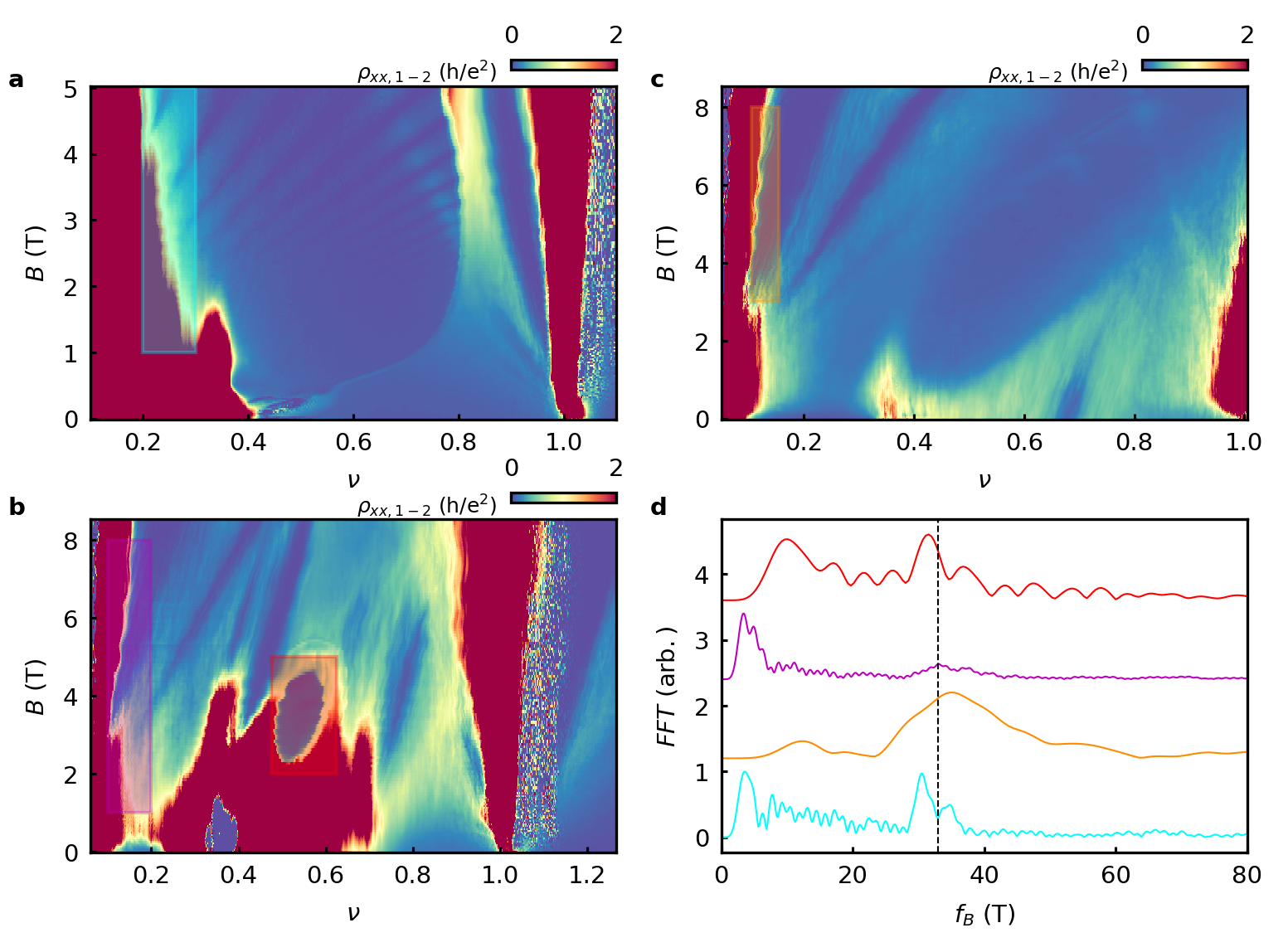} 
\caption{\textbf{Gate-induced quantum oscillations.}
Landau fan diagrams taken at \textbf{a}, $D=0.768\ \rm{V/nm}$, \textbf{b}, $D=0.863\ \rm{V/nm}$, and \textbf{c}, $D=0.82\ \rm{V/nm}$. All of these exhibit oscillation features at constant values of $B$ extending over a wide range of doping.
\textbf{d}, Fast Fourier transform (FFT) of the $\rho_{xx}$ signal with respect to $B$. Each curve is taken in the corresponding colored box in panels \textbf{a}-\textbf{c}. The curves are offset for clarity. Each exhibits a bump above the background at $f_B\approx 33\ \rm{T}$ (dashed black line), although the signal varies substantially depending on the precise values of $\nu$ and $D$. This behavior is similar to that reported in Ref. [1] for regions with large d$R$/d$n$, indicating that the oscillations are features associated with one of the graphite gates rather than intrinsic to the physics of the moir\'e RPG.
}
\label{gate-oscillations}
\end{figure*}

\begin{figure*}
\includegraphics[width=\textwidth]{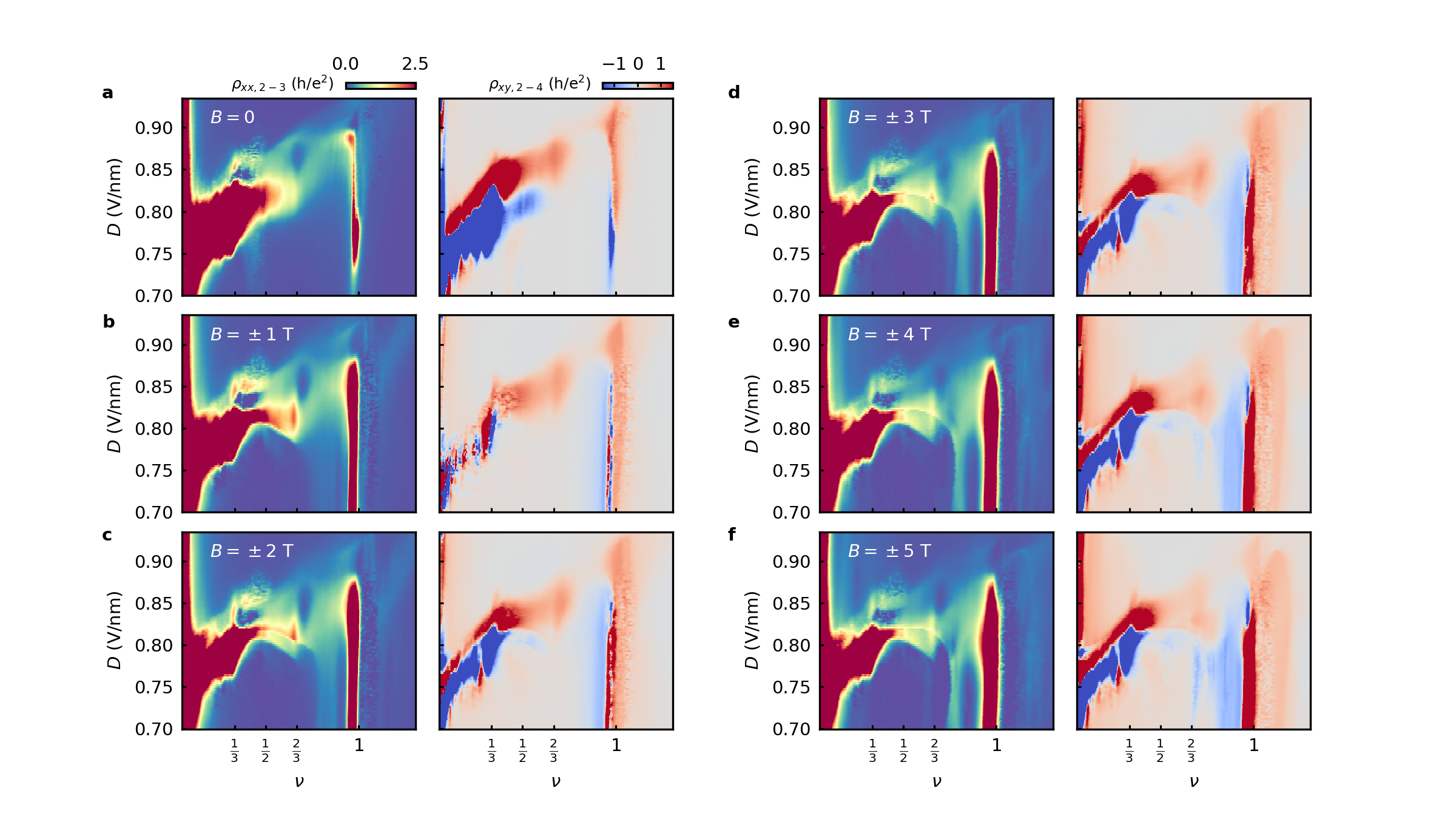} 
\caption{\textbf{Magnetic field evolution of the high $D$ region.}
\textbf{a}, Maps of $\rho_{xx}$ (left) and $\rho_{xy}$ (right) with $B=0$ up to $5$~T. All measurements are taken with $I_{ac}=1\ \rm nA$.
}
\label{AHC_Bfield_dependence}
\end{figure*}

\begin{figure*}
\includegraphics[width=\textwidth]{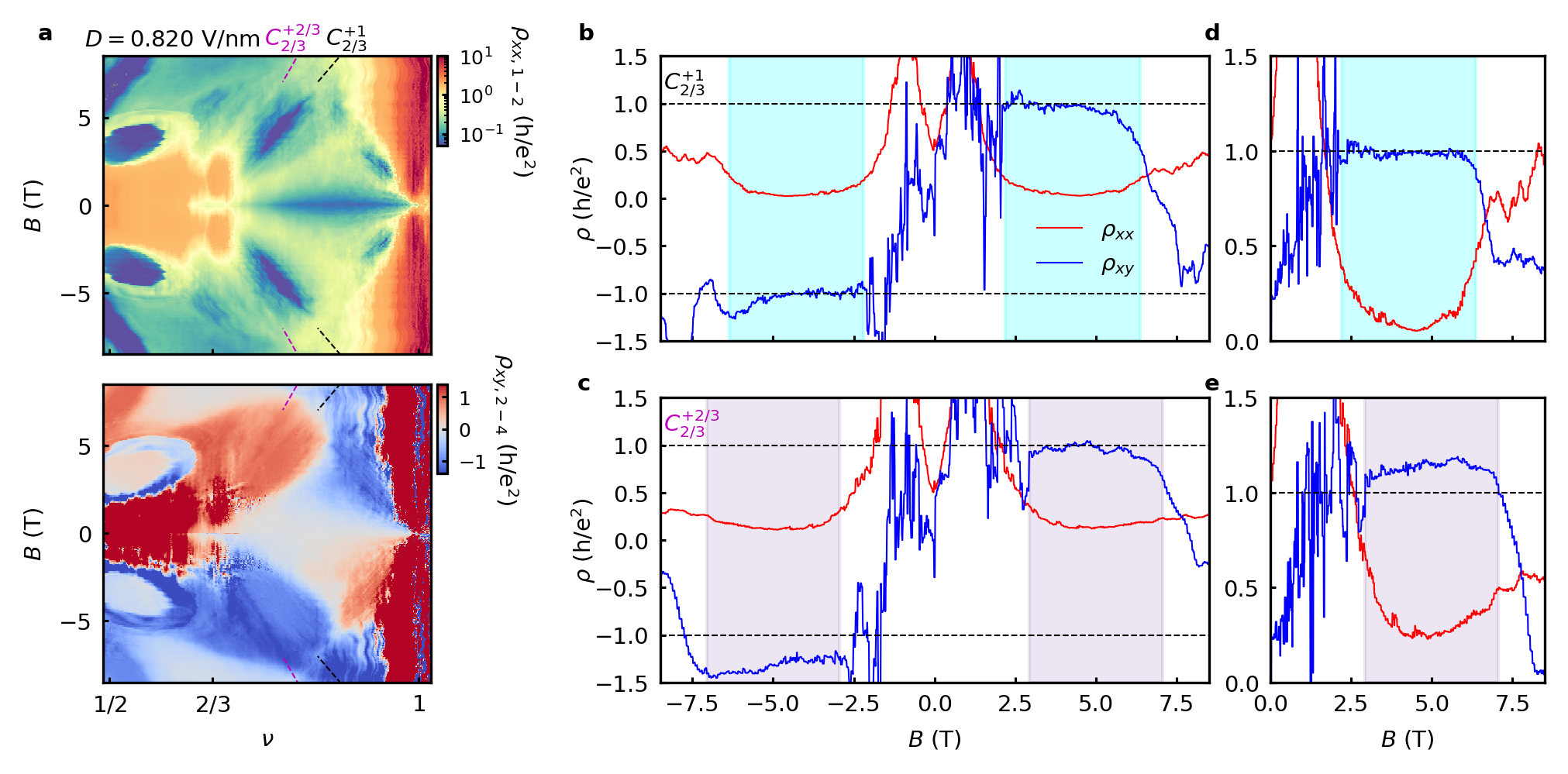} 
\caption{\textbf{Line cuts from unsymmetrized Landau fans.}
\textbf{a}, Landau fan at $D=0.820$~V/nm showing $\rho_{xx}$ without symmetrization (top) and $\rho_{xy}$ without antisymmetrization (bottom).
\textbf{b}, Line cuts along the trajectory of the $C^{+1}_{2/3}$ state. The cyan shaded region is the same magnetic field region identified from the symmetrized data in Fig.~3e (reproduced in \textbf{d}). 
\textbf{c}, Same as \textbf{a}, but for the $C^{+2/3}_{2/3}$ state. The purple shaded region is the same magnetic field region identified from the symmetrized data in Fig.~3f (reproduced in \textbf{e}). We find that there is a significant asymmetry of $\rho_{xy}$ between the positive and negative field values. $\rho_{xy}$ rarely exceeds $h/e^2$ for $B>0$, but approaches the anticipated quantized value of $-3h/2e^2$ for $B<0$.
\textbf{d}, (Anti-) symmetrized curves for comparison with \textbf{b}.
\textbf{e}, (Anti-)symmetrized curves for comparison with \textbf{c}. Because of the non-vanishing $\rho_{xx}$, antisymmetrization is evidently critical for properly assessing the value of $\rho_{xy}$.
\label{current_dependence_bubble}
}
\end{figure*}

\begin{figure*}
\includegraphics[width=6.5in]{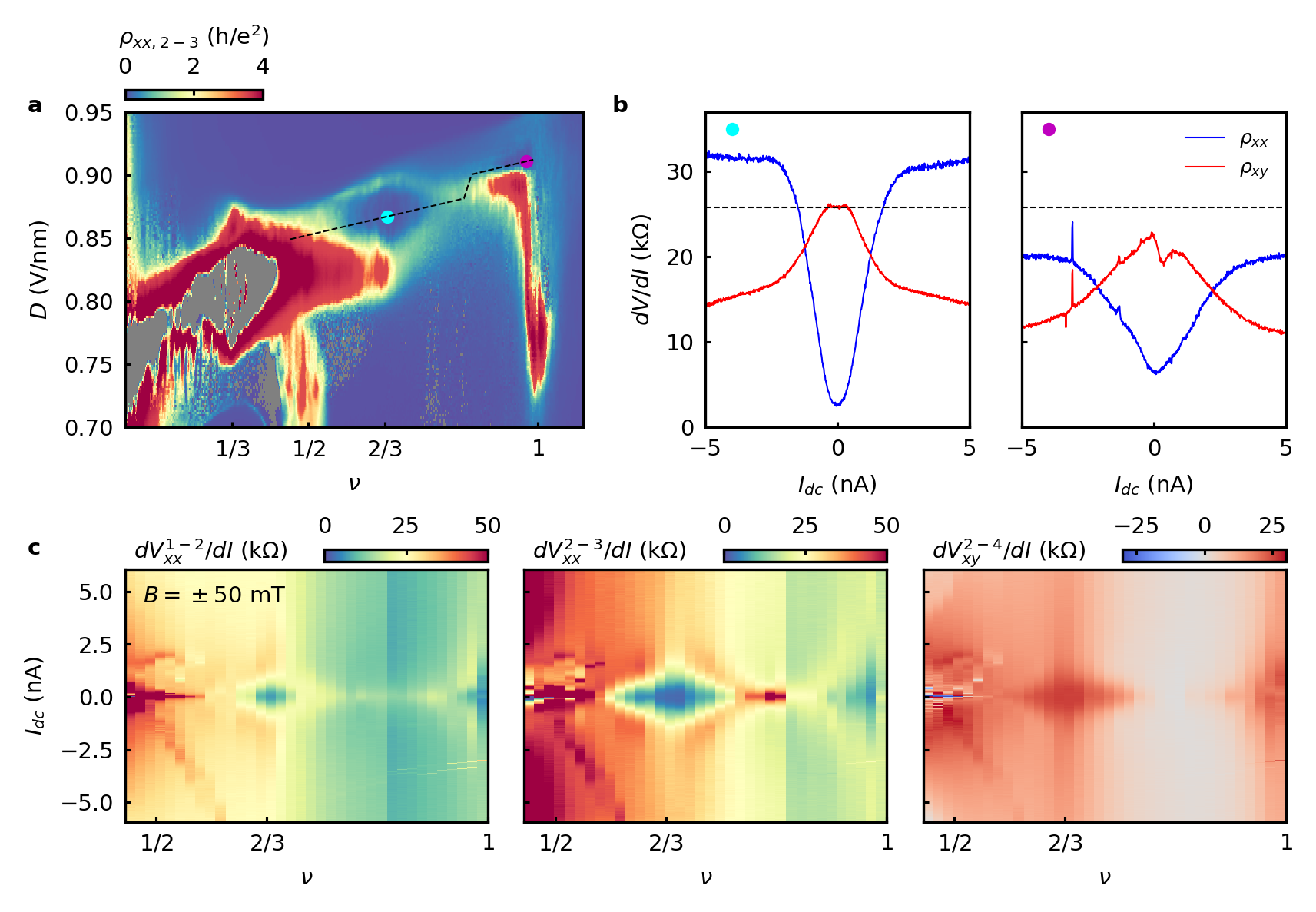} 
\caption{\textbf{Dependence of the TEC states on the bias current.}
\textbf{a}, Map of $\rho_{xx}$ reproduced from Fig.~1e. The purple and blue circle markers indicate the positions of the measurements in \textbf{b}. The black dashed path indicates the path taken for the measurements in \textbf{c}. 
\textbf{b}, Differential resistance, $dV/dI$, measured as a function of $I_{dc}$ with $I_{ac}=14$~pA at the positions of the blue and purple markers in \textbf{a}. At $\nu=2/3$ (blue circle), $\rho_{xy}$ is very nearly quantized to $h/e^2$ and $\rho_{xx}$ exhibits a deep minimum only over a very small range of $I_{dc} \lessapprox 1$~nA. For larger $I_{dc}$, $\rho_{xy}$ is rapidly suppressed and $\rho_{xx}$ spikes to a value in excess of $h/e^2$. Similar behavior is seen at $\nu=1$ (purple dot), although the Chern insulator is not as well developed even for vanishing $I_{dc}$. 
\textbf{c}, Maps of $dV/dI$ as a function of $I_{dc}$ taken along the path of the dashed black path in \textbf{a} (anti-) symmetrized at $|B|=\pm 50\ \rm{mT}$. Each panel corresponds to different contact pairs.
}
\label{sfig:current_dependence}
\end{figure*}

\begin{figure*}
\includegraphics[width=\textwidth]{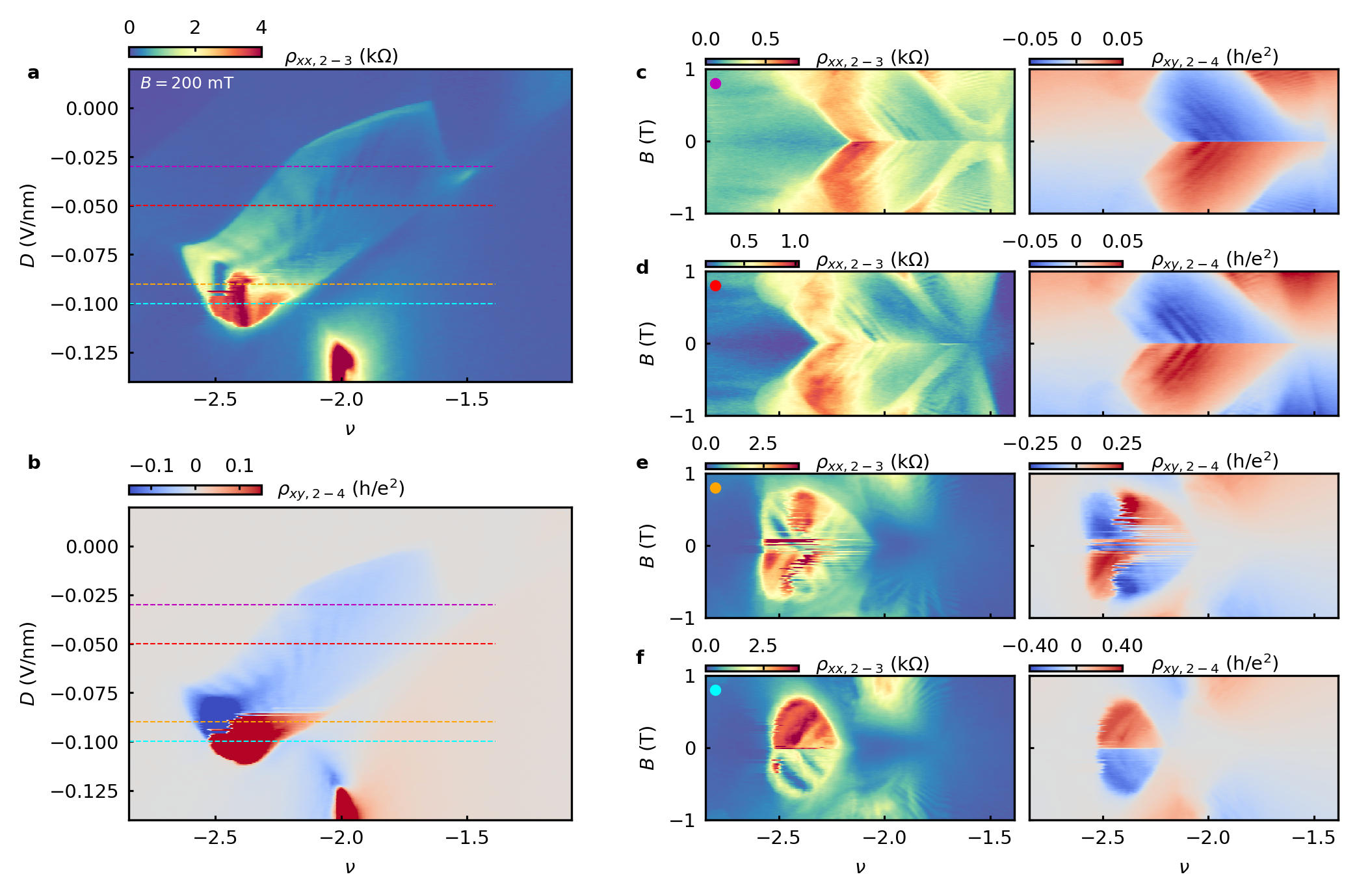} 
\caption{\textbf{Landau fans taken at fixed $D$ through the valley-polarized pocket surrounding $\nu=-2$.}
Maps of \textbf{a}, $\rho_{xx}$, and \textbf{b}, $\rho_{xy}$ at $B=+200\ \rm mT$, reproduced from Fig.~4a. 
\textbf{c}-\textbf{f}, Landau fans taken at constant $D$ (from top to bottom: $-0.03$~V/nm, $-0.05$~V/nm, $-0.90$~V/nm, $-0.10$~V/nm). The colored circles correspond to the dashed lines in \textbf{a}-\textbf{b}. 
All data taken with $I_{ac}=1~\rm nA$.
}
\label{bubble_landauafans}
\end{figure*}

\begin{figure*}
\includegraphics[width=6in]{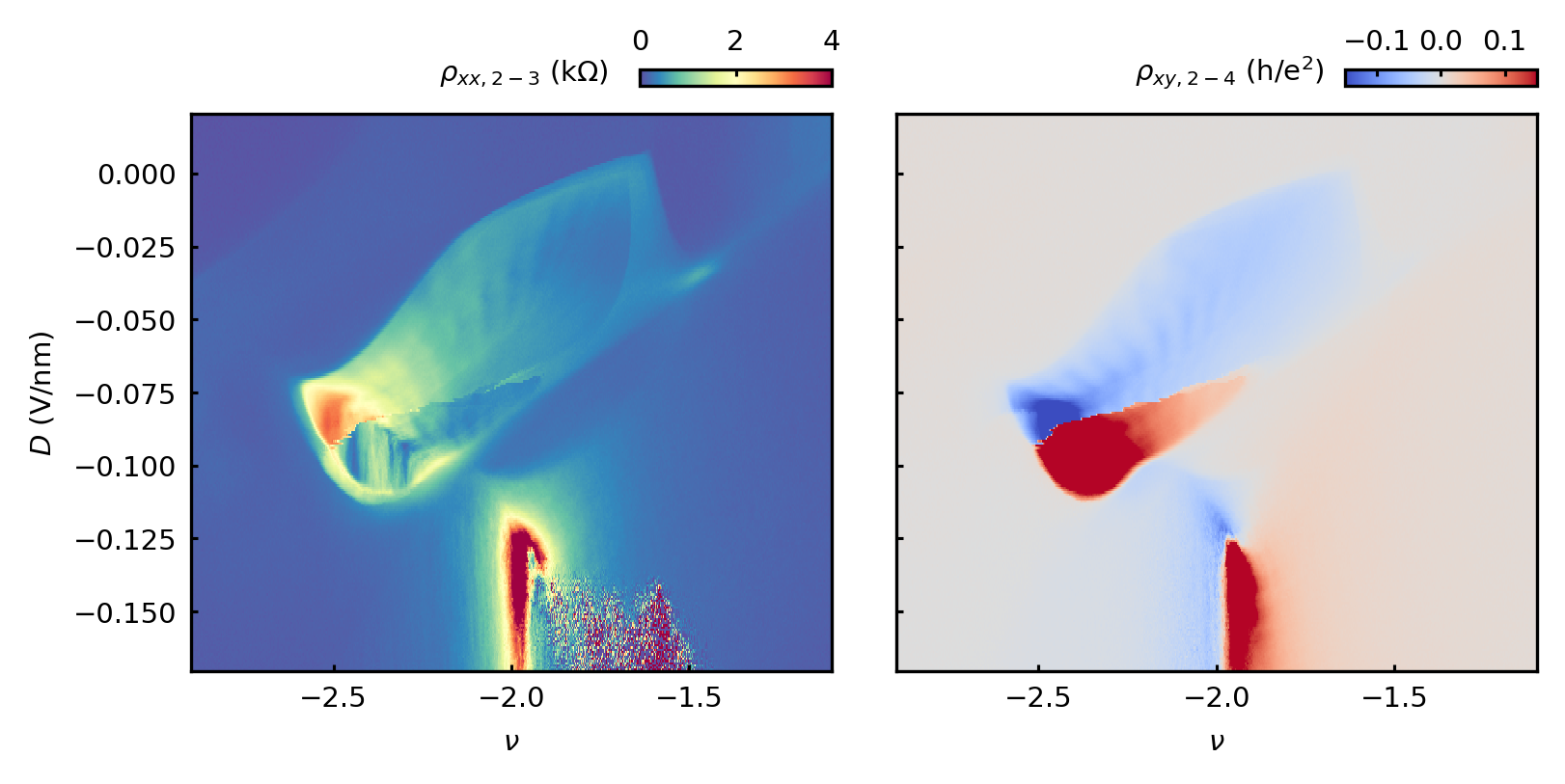} 
\caption{\textbf{Maps of the valley-polarized pocket surrounding $\nu=-2$ taken with lower excitation current.}
Similar maps as those shown in Supplementary Information Fig.~\ref{bubble_landauafans}a-b, but with an ac excitation current of $120$~pA rather than $1$~nA. None of the key features of the maps change as a result of reducing the excitation current (subtle differences between this data and Fig.~\ref{bubble_landauafans}a-b result from sweeping $n$ and $D$ in different directions, which can change the orbital magnetic states, Extended Data Fig. 10d). 
}
\label{current_dependence_bubble}
\end{figure*}

\begin{figure*}
\includegraphics[width=\textwidth]{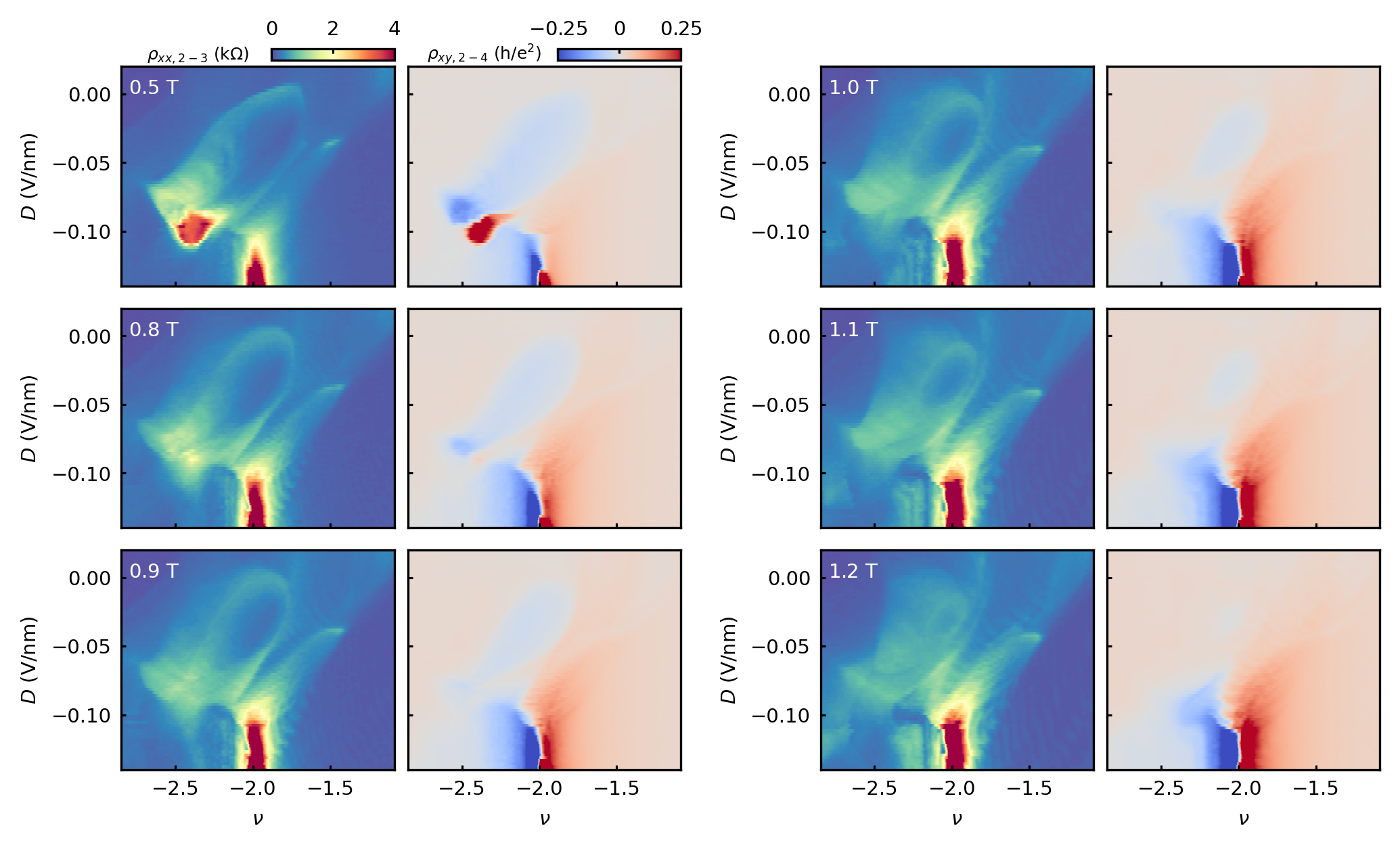} 
\caption{\textbf{Magnetic field--dependence of the valley-polarized pocket surrounding $\nu=-2$.}
Maps of $\rho_{xx}$ and $\rho_{xy}$ taken with various values of the magnetic field as denoted in each panel. The valley-polarized pocket appears to be completely quenched at $B=1.2$~T, consistent with the behavior seen in the Landau fans (e.g., Fig.~4g and Supplementary Information Fig.~\ref{bubble_landauafans}). Data taken with $I_{ac}=1~\rm nA$. Data is not (anti-)symmetrized.
}
\label{bubble_magz}
\end{figure*}

\begin{figure*}
\includegraphics[width=\textwidth]{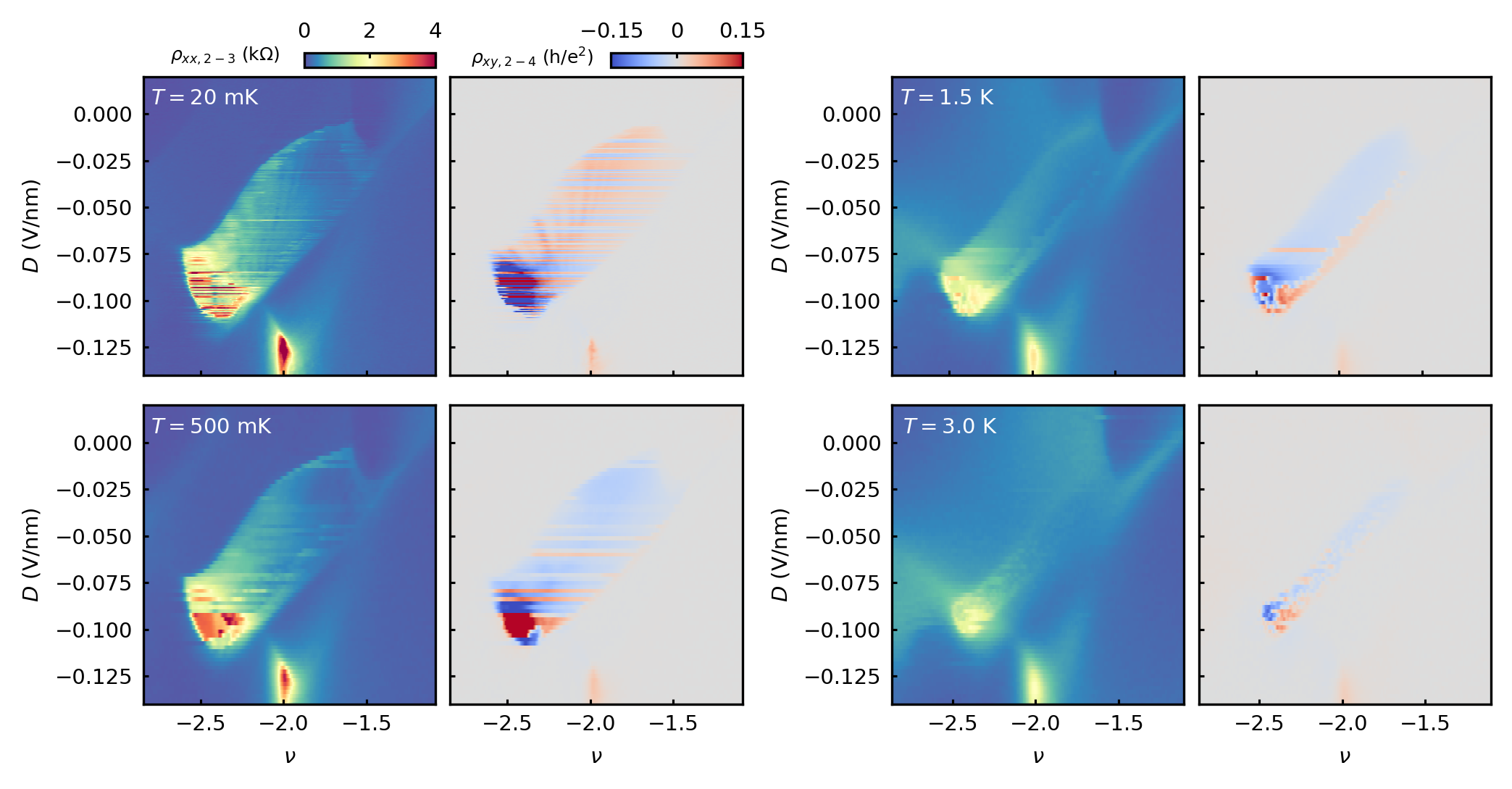} 
\caption{\textbf{Temperature dependence of the valley-polarized pocket surrounding $\nu=-2$.}
Maps of $\rho_{xx}$ and $\rho_{xy}$ taken at various temperatures, as denoted in each panel. The valley-polarized pocket is almost completely suppressed at $T=3$~K. All maps are taken at $B=0$ and with $I_{ac}=1~\rm nA$.
}
\label{bubble_temp}
\end{figure*}

\begin{figure*}
\includegraphics[width=7in]{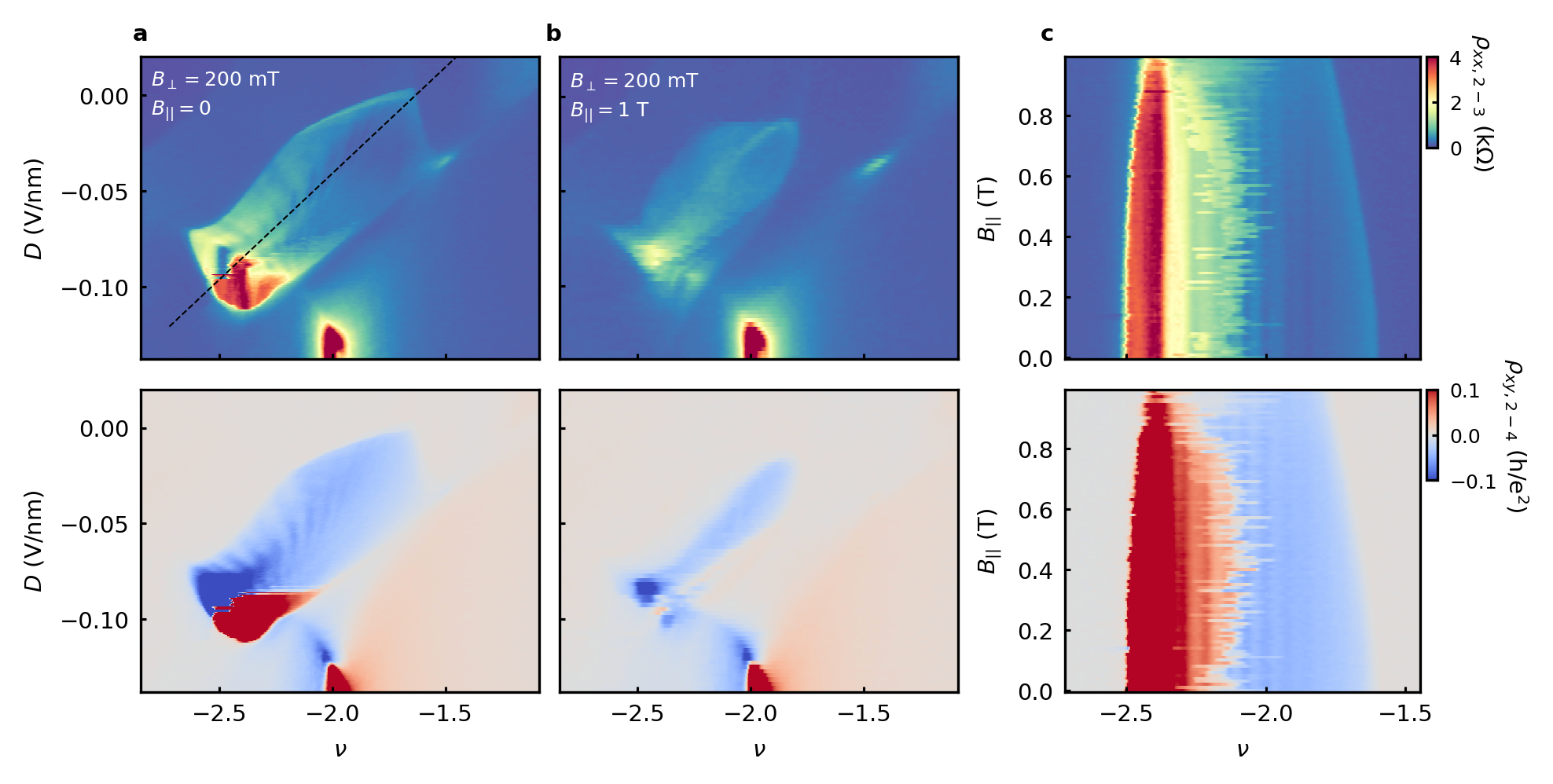} 
\caption{\textbf{In-plane magnetic field dependence of the valley-polarized pocket surrounding $\nu=-2$.}
\textbf{a}, Maps of $\rho_{xx}$ and $\rho_{xy}$ of the correlated states near $\nu=-2$ taken with an out-of-plane field of $B=200$~mT (reproduced from Fig.~4a).
\textbf{b}, The same maps with an additional in-plane magnetic field component, $B_{||}=1\ \rm{T}$. The extent of the valley-polarized pocket in both $n$ and $D$ is reduced as a result of the in-plane field.
\textbf{c}, Maps of $\rho_{xx}$ (top) and $\rho_{xy}$ (bottom) taken as a function of $B_{||}$ with a fixed out-of-plane field of $B=200$~mT. As identified in \textbf{b}, the width of the valley-polarized pocket in $\nu$ is reduced as $B_{||}$ is raised. The oscillations inside the valley-polarized pocket are pinned to fixed values of $\nu$ for all $B_{||}$. Sweeping the doping also induces a flip in the sign of the orbital magnetic state, which also appears to have little dependence on $B_{||}$. Data taken with $I_{ac}=1~\rm nA$. Data is not (anti-)symmetrized.
}
\label{bubble_inplanefield}
\end{figure*}

\begin{figure*}
\includegraphics[width=5.5in]{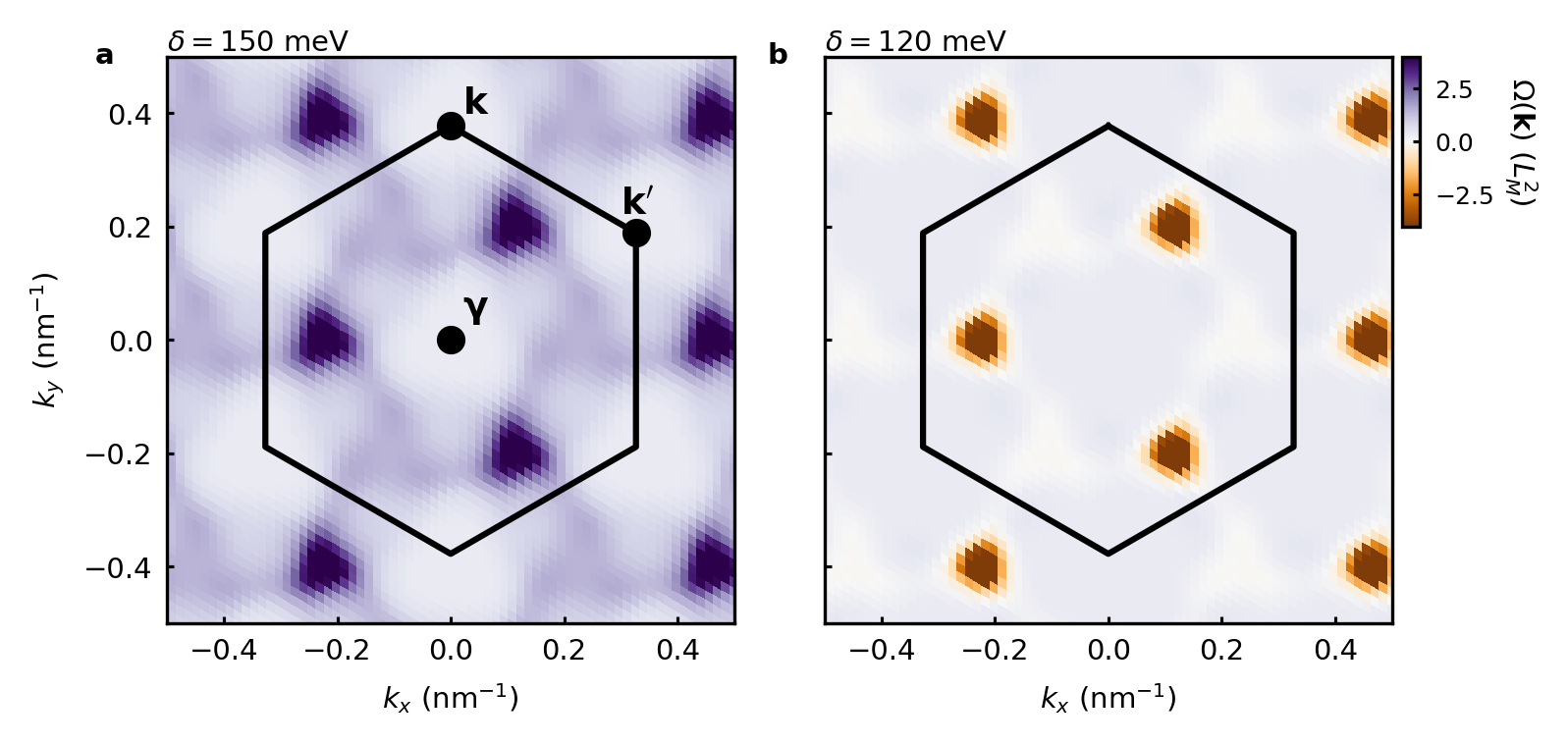} 
\caption{\textbf{Berry curvature distribution of the $\nu=1$ state from Hartree--Fock.}
Berry curvature, $\Omega(\mathbf{k})$, calculated by the Hartree-Fock method at $\nu=1$ (Fig. 2f) for \textbf{a}, $\delta=150\ \mathrm{meV}$ ($\delta>\delta_c$), and, \textbf{b}, $\delta=120\ \mathrm{meV}$ ($\delta<\delta_c$). The black hexagon shows the moir\'e Brillouin zone. The integrated Berry curvature over the entire moir\'e Brillouin zone yields a Chern number of $C=+1$ ($0$) in \textbf{a} (\textbf{b}).
}
\label{efig:HF_Chern}
\end{figure*}

\begin{figure*}
\includegraphics[width=\textwidth]{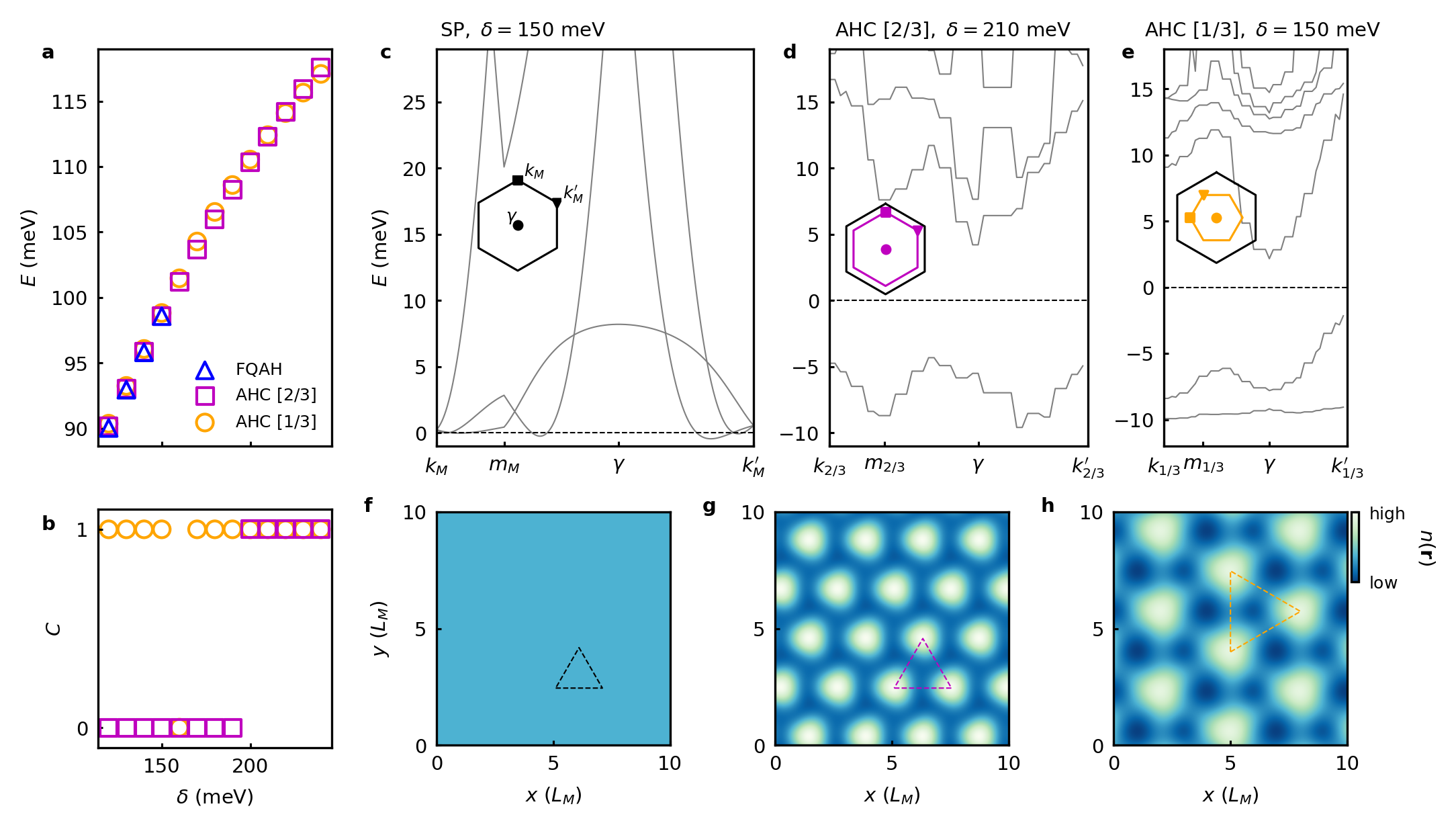} 
\caption{\textbf{Possible ground state orderings at $\nu=2/3$.}
\textbf{a}, Energy of the three possible $\nu=2/3$ ground states as determined by Hartree--Fock calculations in the limit of zero \moire potential as a function of $\delta$ (see Methods for details of the calculations): (I) a FQAH state (II) an AHC state with a period of $L_\mathrm{crystal}=L_M/\sqrt{2/3}$ (AHC [2/3]), and (III) an AHC state with $L_\mathrm{crystal}=L_M/\sqrt{1/3}$ (AHC [1/3]). The energies at each value of $\delta$ are very close for each state, within $0.65~\rm meV$, rendering a meaningful prediction of the ground state ordering impossible. 
\textbf{b}, Corresponding Chern numbers of the two possible AHC states. 
\textbf{c}, Single particle (SP) band structure calculated with the continuum model at $\delta=150\ \rm meV$ (also in the limit of zero \moire potential). The energy axis is taken with respect to the Fermi energy corresponding to $\nu=2/3$ (dashed black line). The inset shows the \moire Brillouin zone (BZ) with high symmetry points labeled. The subscript $M$ denotes the full \moire BZ. 
\textbf{d}, Hartree--Fock band structure for the AHC [2/3] state with $\delta=210~\rm meV$. The dashed line is the Fermi energy, which lies within the gap. The inset denotes the original BZ in black and the reduced BZ of the $L_\mathrm{crystal}=L_M/\sqrt{2/3}$ state in purple. The same high symmetry points are shown as in \textbf{c}, but in the reduced BZ they are denoted by the $2/3$ subscripts. The $\gamma$ point is common between the two BZs.
\textbf{e}, Same as \textbf{d}, but for the AHC [1/3] state. Note that the 1/3 BZ (orange) is rotated by $90^{\circ}$ for commensuration.
\textbf{f}, Calculated real space density for the SP band structure in \textbf{c}. The density is uniform since the calculation is performed in the limit of zero \moire potential. The black dashed triangle denotes the \moire unit cell.
\textbf{g}, Real space density for the HF calculation in \textbf{d}. The dashed purple triangle corresponds to the unit cell with $L_\mathrm{crystal}=L_M/\sqrt{2/3}$. 
\textbf{h}, Real space density for the HF calculation in \textbf{e}. The dashed orange triangle corresponds to the unit cell with $L_\mathrm{crystal}=L_M/\sqrt{1/3}$. \textbf{f-h} are all plotted on the same colorscale. 
}
\label{sfig:hole_crystal}
\end{figure*}

\clearpage

\begin{center}
    \section{References}
\end{center}

\noindent [1] Zhu, J., Li, T., Young, A.F., Shan, J., \& Mark, K.F. Quantum oscillations in two-dimensional insulators induced by graphite gates. \textit{Phys. Rev. Lett.} \textbf{127}, 247702 (2021).